\documentclass[12pt]{article}
\newlength{\abstractwidth}
\usepackage{color}
\usepackage{fancybox}

\newcommand{\bea}{\begin{eqnarray}}
\newcommand{\ena}{\end{eqnarray}}

\newcommand{\ti}{\tilde}

\newcommand{\pd}{\partial}
\newcommand{\MC}{\mathcal}
\newcommand{\MB}{\mathbb}

\usepackage{amsmath, amssymb}	
\begin{document}

\begin{titlepage}
\hspace*{\fill}
\vbox{
\hbox{RUP-21-18}
\hbox{WUCG-21-11}
}
\vskip 1cm

\begin{center}
{\Large{\bf
Rotating particles in AdS:\\ Holography at weak gauge coupling 
and without conformal symmetry}}

\vskip 1cm

Tomotaka Kitamura$^{a}$\footnote{t-kitamura"at"rikkyo.ac.jp},
Shoichiro Miyashita$^{b}$\footnote{s-miyashita"at"aoni.waseda.jp}
and
Yasuhiro Sekino$^{c}$\footnote{ysekino"at"la.takushoku-u.ac.jp}\\

\vskip 0.5cm

$^a${\it Department of Physics, Rikkyo University,
Nishi-Ikebukuro 3-34-1,\\ Toshima,
Tokyo 171-8501, Japan}\\

$^b${\it Department of Physics, Waseda University,
Okubo 3-4-1,\\ Shinjuku,
Tokyo 169-8555, Japan}\\

$^c${\it Department of Liberal Arts and Sciences,
Faculty of Engineering, Takushoku University, \\
Tatemachi 815-1, Hachioji, Tokyo 193-0985, Japan}

\end{center}

\vskip 2cm

\begin{abstract}
We consider gauge/gravity correspondence between 
maximally supersymmetric Yang-Mills theory in ($p+1$) dimensions
and superstring theory
on the near-horizon limit of the D$p$-brane solution.
The string-frame metric is AdS$_{p+1}\times S^{8-p}$ times a Weyl factor,
and there is no conformal symmetry except for $p=3$. 
In a previous paper by one of the present authors,
the free-field result of gauge theory has been reproduced from 
string theory for a particular 
operator which has angular momentum along $S^{8-p}$.  
In this paper, we extend this result to
operators which have angular momenta along AdS$_{p+2}$. 
Our approach is based on a Euclidean formulation proposed by
Dobashi, Shimada and Yoneya and on the ``string bit'' picture.
We first show that the spinning string solution in Lorentzian AdS,
found by Gubser, Klebanov and Polyakov, 
can be recast in a form which connects 
two points on the boundary of Euclidean AdS.
Transition amplitudes of such strings can be interpreted as gauge
theory correlators. 
We study the case of zero gauge coupling by ignoring
interactions among string bits (massless particles in 
ten-dimensional spacetime which constitute a string), 
and show that the free-field results of gauge theory are reproduced.

\end{abstract}
\medskip
\medskip
\medskip
\bigskip
\bigskip

\end{titlepage}
\baselineskip=18pt \setcounter{footnote}{0}


\section{Introduction}

Gauge/gravity correspondence provides concrete realizations 
of the holographic principle~\cite{Holography, tHooft}, a
highly non-trivial proposal which states that quantum gravity 
should be described by  
degrees of freedom localized on the spatial boundary.
Since Maldacena's original proposal of AdS/CFT 
correspondence~\cite{Maldacena}, 
numerous examples of gauge/gravity correspondence have been proposed.
This has led to exciting developments in   
quantum gravity and in strongly coupled quantum systems.
An example of the former would be 
the discovery of the chaos bound~\cite{chaos1}; it was found
through a clever use of the Ryu-Takayanagi formula for 
entanglement entropy~\cite{RT1, RT2} in a two-sided
black hole spacetime which is holographic dual to
the thermofield double state~\cite{chaos2}.
An example of the latter would be the Sakai-Sugimoto models for 
QCD~\cite{SakaiSugimoto1, SakaiSugimoto2}, which have provided 
new viewpoints in terms of strings and branes 
for problems in quantum field theory.

In spite of such developments, gauge/gravity correspondence has not been
proven. In our opinion, there are unclear issues associated to the cases 
without conformal symmetry and/or at weak gauge coupling; clarification 
of these issues could show us a way towards proving 
gauge/gravity correspondence. 

Conformal symmetry has been extremely helpful in formulating
and testing gauge/gravity correspondence. 
Although holographic principle should be a concept independent of
the conformal symmetry, the cases without
conformal symmetry are much less understood than
the cases with conformal symmetry. By correctly formulating 
the correspondence without conformal symmetry, we would be
able to gather important data suggesting how gauge/gravity
correspondence works. 

Gauge/gravity correspondence at weak gauge (or 't Hooft) coupling 
is poorly understood at present. Some people even doubt whether 
correspondence exists in this region, but it is inconceivable that a 
proof of gauge/gravity correspondence is possible
without a clear understanding of this region.
Weak gauge coupling is supposed to correspond to 
weak string tension. In this region, the energy levels of string 
excited states are smaller than the mass scales for the supergravity 
modes (the lowest modes of strings). Thus, supergravity approximation 
is not sufficient, and we will need the string worldsheet 
theory. 

The purpose of this paper is twofold. One purpose is to 
establish a formalism for computing gauge-theory correlators 
using string worldsheet theory in the cases without conformal symmetry. 
The other purpose is to apply this formalism to the case of zero
gauge coupling, with an additional assumption of ``string bit.'' 
We will reproduce the free-field result of gauge 
theory correlators from string theory, extending the result
obtained in a previous paper by one of the present 
authors~\cite{Sekino} for a particular operator
to the general class of operators.

In Sec.~\ref{preliminaries} below, we will remind the reader of
basic facts about gauge/gravity correspondence. This part 
can be skipped if it is not necessary for the reader. 
In Sec.~\ref{Dp-branes}, we will introduce the concrete example studied 
in this paper, namely the gauge/gravity correspondence associated 
with D$p$-branes. This part is a review of old work, but 
may contain subjects that are not widely known. Then, 
in Sec.~\ref{subject}, we will describe the aim of the present work. 

\subsection{Conformal and non-conformal}
\label{preliminaries}
Conformal symmetry has played important roles in 
formulating and testing gauge/gravity correspondence, such
as the following:
(1)~The first hint for the 
equivalence of two completely different theories was that
the symmetry on both sides match~\cite{Maldacena}: 
The isometry group of $(d+1)$-dimensional Anti-de-Sitter (AdS) space 
and the conformal group in $d$ dimensions are isomorphic.
(2)~The dictionary between the 
bulk fields and the boundary operators, namely, the coupling between 
them in the Gubser-Klebanov-Polyakov-Witten (GKPW) 
prescription~\cite{WittenAdS, GKP}, is determined from the requirement that 
they should belong to the same 
representations of the (super) conformal group (see e.g., \cite{Ferrara}). 
(3)~In theories with superconformal symmetry, there are powerful 
non-renormalization theorems. For instance, the scaling dimensions of the 
so-called BPS operators, which correspond to supergravity (SUGRA) modes,
 are not renormalized from their free-field values. The fact that the
gauge-theory correlators calculated by the GKPW prescription using the 
tree-level SUGRA have the free-field scaling dimensions 
provides an important consistency check of AdS/CFT correspondence.
(4)~There have been highly non-trivial tests of AdS/CFT correspondence
from calculations of quantities interpolating between weak and 
strong couplings. In such analyses,
conformal symmetry played essential technical roles.
Examples include the calculation of the expectation values of 
Wilson loops by using conformal transformations of the shape 
of the loops (see e.g., \cite{Semenoff}) and the calculation of 
cusp anomalous dimensions using integrability 
(see e.g., \cite{Eden}). 

In theories without conformal symmetry, we do not have 
these (at least not obviously),
and it is not straightforward to study gauge/gravity correspondence. 

In fact, there have been many examples of gauge/gravity correspondence 
without conformal symmetry.
However, most of them are associated to theories with conformal symmetry 
(though they involve highly non-trivial and interesting ideas),
in the sense that they are continuous deformations of theories
with conformal symmetry, and the symmetry is restored in some limit.
Examples of such interesting theories include: (a) Attempts at 
holographic duals of QCD-like theories, starting from the work by
Witten~\cite{WittenThermal}, by compactifying spatial directions 
in the conformally invariant 6D theory; (b) Studies of RG flows by 
constructing geometries which interpolate between two AdS 
regions~\cite{PilchWarner, KlebanovStrassler, PolchinskiStrassler}; 
(c) Applications of gauge/gravity correspondence to nuclear or
condensed matter physics (see e.g., \cite{LiuReview, HartnollReview} 
for reviews), in which attention is focused on  
the vicinity of the quantum phase transition points
at which conformal symmetry is realized. 

There are very few examples of gauge/gravity correspondence 
that do not have conformal symmetry from the outset. One of such 
examples is the one described below. Through the study of such a
theory, we hope to develop the formalism and techniques applicable
to non-conformal cases in general.

\subsection{Gauge/gravity correspondence for D$p$-branes}
\label{Dp-branes}
In this paper, we will study the gauge/gravity correspondence 
between maximally supersymmetric $SU(N)$
Yang-Mills theories in ($p$+1) dimensions and superstring theories 
on the near-horizon limit of the D$p$-brane solutions, proposed
by Itzhaki, Maldacena, Sonnenschein and Yankielowicz~\cite{IMSY}. 
The former theory is an open-string description on the worldvolume 
of D$p$-branes, and the latter is a closed string description 
treating the D$p$-branes as classical solutions in supergravity. 
The $p=3$ case is conformally invariant, and is the most
typical example of AdS/CFT correspondence~\cite{Maldacena}. 
For $p\neq 3$, the gauge coupling 
has non-zero dimension, and the theory is not conformally invariant. 
Without conformal symmetry, the analysis for $p\neq 3$ is not easy, 
but the correspondence is as well-motivated as in the $p=3$ case.  
We expect the $p\neq 3$ cases to yield useful data about
the mechanism of gauge/gravity correspondence. 
In this paper, we will study this example of gauge/gravity
correspondence. 

Supergravity analysis based on the GKPW prescription 
has been performed on the near-horizon D$p$-brane 
background~\cite{SekinoYoneya, SekinoSupercurrents, SkenderisDp},
and it has been found that the operators in the maximally 
supersymmetric Yang-Mills theories in $(p+1)$ dimensions
which correspond to SUGRA modes have power-law correlators, 
even though there is no conformal symmetry
for $p\neq 3$. This is the result supposed
to be valid at strong 't Hooft coupling.
The power is in general a fractional number, different from 
the free-field value. This power law has not been understood analytically
in gauge theory, but for some operators in $p=0$, 
it has been confirmed to a high precision by the 
Monte-Carlo simulation in gauge theory~\cite{HNSY1, HNSY2}. 

A problem with the cases without conformal symmetry has been
that it is not clear how to perform bulk analysis based 
on the string worldsheet. For theories with conformal symmetry, 
one can identify the energy with respect to the global time
in AdS with the scaling dimension of the corresponding 
operator in gauge theory. This identification is based on the 
isomorphism of the symmetry groups, and cannot be applied to
the case without conformal symmetry.  
Although the superstring action on AdS has complicated 
non-linear interactions and is difficult to solve, there has been 
significant progress based on semi-classical approximations, 
especially\footnote{%
Also, in other cases such as AdS$_3$/CFT$_2$~\cite{MaldacenaOoguri1, 
MaldacenaOoguri2,
MaldacenaOoguri3} and D1-D5 system (see e.g., \cite{D1D5}), 
detailed worldsheet analyses have been 
performed.} 
in the case of AdS$_5\times S^5$:
(a) In the limit of large angular momenta along $S^5$, Berenstein, Maldacena
and Nastase (BMN) found the spectrum of quadratic fluctuations around
a point-like classical configuration of the string moving along $S^5$, 
and obtained the scaling dimensions of the corresponding operators
(the so-called BMN operators). This includes higher string excitations, 
not only the supergravity modes~\cite{BMN}. 
(b) In the limit of large angular momenta along AdS$_5$, Gubser, 
Klebanov and Polyakov (GKP) identified classical solutions of strings
with such momenta, which are folded and spinning in AdS,  
and obtained the scaling dimensions of the corresponding operators 
in gauge theories. This result is expected to capture the
properties of non-supersymmetric theory as well~\cite{GKP2}. 

To the best of our knowledge, the only formalism for 
the worldsheet analysis applicable to the non-conformal cases 
are the one proposed by Dobashi, Shimada and Yoneya (DSY)~\cite{DSY}. 
In the original paper of DSY, the BMN operators 
in the conformally invariant $p=3$ case
were studied. They consider Euclidean AdS in order to make 
contact with the GKPW prescription, which is formulated in the 
Euclidean background. There is a geodesic in Euclidean AdS 
which connects two points on the boundary. By performing a semi-classical 
approximation (similar to the one performed by BMN) of superstring along this 
geodesic, one can compute transition amplitudes of the Euclidean string. 
The amplitude is interpreted
as the correlation function of the corresponding BMN operator. 
The results thus obtained are consistent
with the ones obtained by BMN~\cite{BMN}; furthermore, some puzzles 
have been solved (see \cite{DSY, DobashiYoneya1, Dobashi}). 
This formalism does not use conformal symmetry and is based on 
an intuitively clear relation between the bulk and boundary, thus
it is applicable to the non-conformal cases.
Asano, Yoneya and one of the present authors have applied 
this to the study of the BMN operators for general 
$p$ ($0\le p\le 4$)~\cite{ASY, AsanoSekino}. 
When applied to the SUGRA modes, this method gives the result 
consistent with the one obtained by the GKPW prescription.
The bulk analyses described above are supposed to give results
for gauge theory in the large $N$ limit 
with strong 't Hooft coupling. 

\subsection{Aim of the present work}
\label{subject}

As reviewed above, it is essentially understood how to compute
correlation functions at strong 't Hooft coupling from the bulk
(using supergravity or string theory),
including the cases of $p\neq 3$ which are not conformally invariant.
On the other hand, it has not been known how to obtain the free-field
results of gauge theory from string theory except for
the BPS operators for $p=3$, whose scaling dimensions 
are protected.

In a previous paper by one of the present authors~\cite{Sekino}, 
the BMN operators in the ($p+1$)-dimensional super Yang-Mills 
theory have been studied at zero gauge coupling in string theory.
This is based on the ``string bit'' picture, summarized
as follows: Consider single trace operators, ${\rm Tr}(Z^J)$, 
where $Z$ is a complex combination of two of the
$(9-p)$ scalar fields in gauge theory, say, $Z=\phi_8+i\phi_9$,
defined in the manner similar to the $p=3$ case,
following BMN. We will call this a BMN operator. 
(It might be common to use this terminology only when
$J$ is large, but here we will continue calling so
when $J$ is not necessarily large.) This operator corresponds 
to a string state with $J$ units of angular momentum along
$S^{8-p}$. We assume the spatial direction of the worldsheet 
is discretized into $J$ bits, each of which has a single unit
of angular momentum. 

The reasons for considering the string bits are twofold: 
First, on the gauge-theory side, if one wants to
represent the string worldsheet by the cyclic sequence of
the fields inside the trace, 
one can only represent $J$ sites. String bit picture is 
mentioned in the original paper by BMN~\cite{BMN}. It also
appears in recent papers such as \cite{Minwalla} and
\cite{Gaberdiel} in the contexts related but somewhat 
different from ours\footnote{%
We thank Yasuaki Hikida for pointing out the former
reference to us.}.

In our opinion, there is another reason for considering 
string bits, on the string theory side:
To represent a state with non-zero angular momentum, 
one usually inserts one or more creation operators
which have appropriate angular momentum. Those operators
are the Fourier modes with respect to the spatial direction
on the worldsheet.
In other words, they are obtained by smearing local
operators on the worldsheet. 
Before smearing, each operator inserted on the worldsheet can
be regarded as a particle (bit) with a unit angular momentum,
interacting with other bits via strings connecting them. 
At weak gauge coupling limit, the unit of angular momentum is 
large\footnote{%
In this limit in which the string length is much larger than
the AdS radius, stretched strings within a patch with the AdS 
size cannot be ignored. The string bit picture may provide one
approach for understanding ``sub-AdS locality,'' 
known to be a difficult problem in gauge/gravity 
correspondence~\cite{SusskindWitten}.},
and the interactions among the bits are weak. Thus, it
would be appropriate to treat these bits as discrete objects.

In the previous paper~\cite{Sekino}, the case of zero coupling 
was studied by ignoring the interactions among the bits. 
Then, following the DSY formalism~\cite{DSY, ASY} 
mentioned in the last subsection, 
the amplitude for the collection of bits was computed, and 
interpreted as a gauge-theory correlator. By properly taking into 
account the zero-point energies due to the fluctuations,  
the free-field result of the $(p+1)$-dimensional field theory
was reproduced. 

The result of \cite{Sekino} was obtained for one particular operator. In this
paper we would like to show that the free-field result can be obtained 
for general operators. We consider operators which have angular 
momentum along AdS, typically of the form $\sum_{i}{\rm Tr}
(\phi_i D^S \phi_i)$,
where $\phi_i$ are scalar fields in the $(p+1)$-dimensional gauge 
theory, and $D$ denotes complex combination of gauge covariant derivatives
in two directions, say, $D=D_1+iD_2$. 
This type of operators are called Gubser, Klebanov and
Polyakov (GKP) operators~\cite{GKP2}.
(It might be common to use this terminology only when
$S$ is large, and angular momentum 
along the $S^{8-p}$ directions is zero, but we will use this terminology 
loosely, and use it whenever an operator has non-zero angular momentum 
along AdS$_{p+2}$.) At strong 't Hooft coupling for $p=3$, 
the GKP operators are described by strings which are folded and spinning
in AdS, as found in a seminal paper by GKP~\cite{GKP2}. 
In the present paper we first
clarify how to describe such operators in the bulk for $p\neq 3$
(in which we cannot rely on the identification between the energy 
in terms of global time and the scaling dimension).
Our approach is an extension of the Euclidean formulation of
DSY, in which an angle along which the string is moving
is taken to be imaginary, in order to keep the
angular momentum real in the Euclidean setting. In DSY, only 
one angle was imaginary, but in our work, two of
the angles are taken to be imaginary.
In this formulation, 
we will find a solution in which a string worldsheet connects 
two points on the boundary; the center of the string follows a 
trajectory in AdS similar to the one for the BMN operators.

Then, we consider zero gauge coupling in this framework. As in the 
previous paper~\cite{Sekino}, we assume a bit has a unit angular momentum
along $S^{8-p}$. Calculating the amplitudes by ignoring the
interactions among the bits, we will find the free-field results 
both for $p=3$ and for $p\neq 3$. (For $p=3$, we will also present
an analysis for zero gauge coupling in the Lorentzian formulation.)
We regard these results to be indications for the validity of 
the approach initiated in~\cite{Sekino}.

We should mention that our result is based on two assumptions
which have not been proven at present. 
One is that we assume the near-horizon
D$p$-brane background does not receive $\alpha'$ corrections
for general $p$. The $p=3$ case is believed to be
$\alpha'$ exact~\cite{Banks, Kallosh}, 
but for $p\neq 3$ it is not known, to the
best of our knowledge.
The other assumption is that when we compute the zero-point 
energy of a single bit, we take into account the fluctuations
up to the quadratic order around the classical trajectory. 
For $p=3$, contributions from the bosonic and fermionic fluctuations
cancel with each other because there is effectively a worldsheet 
supersymmetry~\cite{Metsaev1, Metsaev2, Cvetic, AsanoSekino}.
But for $p\neq 3$, we do not have a clear explanation why 
it is enough to stop at the quadratic order. Since we are 
not considering large angular momenta (because we are
considering a single bit), there is no small parameter, so 
higher order terms could in principle contribute. 
One way to estimate the order would be as follows: a bit can be 
regarded as a massive particle in AdS$_{p+2}$; its mass is 
given by the angular momentum on $S^{8-p}$ and is of order 
$m\sim 1/L$, where $L$ is the radius of AdS$_{p+2}$ and $S^{8-p}$. 
Thus, there is an expansion parameter $1/m\sim L$, but
each order is associated with the corresponding factor 
of $1/L$, since the expansion along the geodesic in 
AdS$_{p+2}$ is essentially an expansion in curvature. 
In this counting, each term is of order unity. (This is also
clear from the fact that the action, 
$m\int d\tau \sqrt{g_{\mu\nu}\dot{x}^{\mu}\dot{x}^{\nu}}$,
of massive particle on AdS$_{p+2}$
has no dependence on $L$, since $m\sim 1/L$, and 
$\sqrt{g_{\mu\nu}\dot{x}^{\mu}\dot{x}^{\nu}}\sim L$.)

\subsection{Organization of this paper}
This paper is organized as follows. 
In Section 2, we will briefly review gauge/gravity 
correspondence for D$p$-branes. Basic 
features of the near-horizon limit of the D$p$-brane solution
will be described. 
In Section 3, we will consider the case of strong gauge coupling.
We will first review the Lorentzian solution found by Gubser, Klebanov and 
Polyakov, then, we rotate some coordinates to the imaginary
directions, and obtain a string configuration which connects 
two points on the boundary. We will show that the string amplitude
can be interpreted as the gauge-theory correlator. 
Our analysis here is based on the $p=3$ case, but this formalism
is applicable to the cases without conformal symmetry. 
In Section 4, we consider weak gauge coupling. 
We study the zero coupling case
by considering the string bits without interactions
among them. We first describe the analysis for $p=3$ using global time
in Lorentzian AdS. 
From the energy of the rotating particle, we obtain the free-field 
result in gauge theory. We then study the case of general $p$. We 
compute the amplitude for particles moving along a trajectory connecting 
two points on the boundary, and show that this reproduces the
free-field result in gauge theory.  
In Section 5, we conclude, and mention directions 
for future research. 

\section{Gauge/gravity correspondence}
\label{GaugeGravity}
Gauge/gravity correspondence associated to the D$p$-branes has been
proposed by Itzhaki, Maldacena, Sonnenschein and 
Yankielowicz~\cite{IMSY}. We will assume $0\le p\le 4$.
Here we will review only the basic facts, and refer the readers to
the original papers for details\footnote{%
See also~\cite{Jevicki1, Jevicki2, Jevicki3} 
for ``generalized conformal symmetry,'' which 
motivated the following analyses.}:
for supergravity analysis based on the
GKPW prescription, see~\cite{SekinoYoneya, 
SekinoSupercurrents, SkenderisDp}; for 
worldsheet analysis, see~\cite{ASY, AsanoSekino, Asano}; 
for the tests by Monte Carlo simulations in gauge theory, 
see~\cite{HNSY1, HNSY2};
for the correspondence at weak gauge coupling, see~\cite{Sekino}.

The metric and the dilaton for 
the zero-temperature D$p$-brane solution in the string frame is given 
by\footnote{%
Time coordinate $t$ in this equation corresponds to time in the 
Poincar\'{e} coordinates, which will be denoted as $t_{\rm P}$
in Section~3. Time for global coordinates will be
denoted $t$ there.}
\begin{eqnarray}
ds^{2}&=&H^{-1/2}\left(-dt^{2}+d x_{a}^{2}\right)
+H^{1/2}\left(dr^{2}+r^{2}d\Omega_{8-p}^{2}\right),
\nonumber\\
e^{\phi}&=& g_{s}H^{\frac{3-p}{4}},
\quad H = 1+{q\over r^{7-p}} 
\label{eq:nhDp}
\end{eqnarray}
where $a=1,\ldots, p$ and 
$q=\tilde{c}_{p}g_{s}N\ell_{s}^{7-p}$ with 
$\tilde{c}_{p}=2^{6-p}\pi^{(5-p)/2}\Gamma{(7-p)/2}$. 
The integer $N$ denotes the number of the D$p$-branes,
$\ell_s$ is the string length, and $g_s$ is related to 
the Yang-Mills coupling by $g^2_{\rm YM}=(2\pi)^{p-2}g_s\ell_s^{p-3}$.
We consider the near-horizon limit $r\ll q^{1/(7-p)}$,
and take 
$H\to {q/r^{7-p}}$.

For $p=3$, the near-horizon geometry is AdS$_5\times S^5$. For
$p\neq 3$, it is related to AdS$_{p+2}\times S^{8-p}$
by a Weyl transformation\footnote{%
For $p=5$, the background is Weyl equivalent to a linear dilaton
background.%
}: 
\begin{equation}
ds^2 =H^{1/2}r^2 
\left[
\left({2\over 5-p}\right)^{2}\left( dt^{2}+dx_{a}^{2}+dz^{2}\over z^{2}\right)
+d\Omega_{8-p}^2\right].
\label{eq:Weyl}
\end{equation}
The radial variable $z$ in the Poincar\'{e} coordinates
for AdS$_{p+2}$ is defined by
\begin{equation}
z={2\over 5-p}(g_sN)^{1/2}l_s^{(7-p)/2}r^{-(5-p)/2}={2\over 5-p}H^{1/2}r.
\end{equation}
For $p=3$, the Weyl factor $H^{1/2}r^2$ is constant.
For $p\neq 3$, there is no AdS isometry, since the Weyl factor, dilaton,
and the gauge fields do not have such symmetry. 
The radius of AdS$_{p+2}$ is $2\over 5-p$ times the radius of $S^{8-p}$,
as we see from the relative factor between the first and the second
terms in \eqref{eq:Weyl}.

To the best of our knowledge, it is not known whether there are 
$\alpha'$ corrections to the background \eqref{eq:Weyl} for general $p$.
We assume there are no corrections, and continue using this background 
when the curvature is strong.

The string coupling $e^{\phi}$ and the curvature of the background
depend on the position for $p\neq 3$. String coupling is weak,
$e^{\phi}\ll 1$, away from the center $r/\ell_s\gg N^{{-4\over (3-p)(7-p)}}
(g_sN)^{1/(3-p)}$ for $p<3$, and away from the boundary
$r/\ell_s\ll N^{{4\over (p-3)(7-p)}} (g_sN)^{-1/(p-3)}$ for $p>3$.
To study the curvature, it would be helpful
to consider an effective curvature radius $\tilde{L}(r)$ given by the Weyl
factor in \eqref{eq:Weyl}, which can be written as
\begin{equation}
\tilde{L}^2(r)\equiv H^{1/2}r^2\propto (g_s N\ell_s^4)^{1/2}
\left({\ell_s\over r}\right)^{(3-p)/ 2}.
\end{equation}
The curvature is weak relative to the string scale,
$\tilde{L}(r)\gg \ell_s$, away from the boundary 
$r/\ell_s\ll (g_sN)^{1/(3-p)}$ for $p<3$, and 
away from the center $r/\ell_s\gg (g_sN)^{-1/(p-3)}$
for $p>3$.
If one wants to use the tree-level supergravity approximation, 
we will need these conditions to be satisfied in a large
portion of the near horizon region. 
This is achieved if we take $N\to \infty$ with $g_s N$ 
(i.e., the 't Hooft coupling $g^2_{\rm YM}N$) fixed but 
large~\cite{SekinoYoneya}\footnote{%
The singularity at the center $r=0$ causes no problem in the GKPW
prescription, since we take the wave functions that decay 
exponentially at the center~\cite{GKP}.}. 
In this paper, we will always ignore string loops. This corresponds
to taking $N\to \infty$ in gauge theory. In Sec.~\ref{WeakCoupling}, 
we will consider weak gauge coupling $g_s\to 0$ (or weak 't Hooft 
coupling $g_sN\to 0$), which corresponds
to strong curvature $\tilde{L}(r)/\ell_s\to 0$. We will study this
region using string theory based on the string bit picture, 
not relying on supergravity approximation.
When gauge coupling is zero, the effective string tension, which is 
of order $\tilde{L}^2(r)/\ell^2_s$, is zero.  

Gauge theory which corresponds to superstring theory
on the above background is the
maximally supersymmetric $SU(N)$ Yang-Mills theory in $(p+1)$ dimensions.
This theory is obtained by dimensional reduction from the 
super Yang-Mills theory in (9+1) dimensions, and
has ($9-p$) scalar fields, $\phi_{p+1}\ldots, \phi_{9}$,
in addition to the gauge field in $(p+1)$ dimensions. There is 
$SO(9-p)$ global symmetry which rotates the scalar fields
among themselves. We will consider a complex combination
of two scalar fields, say, $Z\equiv \phi_8+i\phi_9$, and
also a complex combination of gauge covariant derivatives
in two directions, say, $D\equiv D_{1}+iD_{2}$. 

\section{Spinning strings (strong gauge coupling)}
\label{StrongCoupling}
The purpose of this section is to establish a method for 
calculating correlators of operators which have angular 
momentum along AdS$_{p+2}$ from string theory
that is
applicable to the $p\neq 3$ cases which are not conformally 
invariant. 
Our method is an extension of the prescription proposed by 
Dobashi, Shimada and Yoneya, which has so far been 
formulated only for the BMN operators (those with angular 
momenta along $S^{8-p}$). In this formulation, 
we will consider Euclidean AdS. 
In the limit of strong 't Hooft coupling, correlation functions
of operators of the form $\sum_{i}{\rm Tr}(\phi_i D^S \phi_i)$ 
can be calculated by evaluating the action of classical
string solutions. For $p=3$, the string 
solution is given from the one obtained by GKP~\cite{GKP2}
by rotating some coordinates to imaginary values. It connects
two points on the boundary. We will defer the study of classical
solutions for $p\neq 3$ to future work, but this formalism
is applicable to the cases without conformal symmetry.

In Sec.~\ref{LorentzianString}, we review GKP's analysis
performed in Lorentzian AdS for $p=3$. 
Then, in Sec.~\ref{EuclideanString}, we obtain a solution which 
connects two points on the boundary in the Euclidean formulation.
We will see that the Euclidean amplitude for this string
can be written in the form of gauge-theory correlator.

\subsection{Lorentzian signature}
\label{LorentzianString}
We write the AdS$_5\times S^5$ using 
the global coordinates for AdS\footnote{%
Time coordinate $t$ without a subscript will denote
time in global coordinates in this section and Sec.~4.1.},
\begin{align}
ds^2& = L^2 \left\{ 
- \cosh^2\rho\,  dt^2+d\rho^2 + \sinh^2 \rho 
\left(\cos^2\theta\,d\psi^2
+d\theta^2 +\sin^2 \theta d\phi^2
\right)\right.\nonumber\\
&\qquad \quad \left.
+\cos^2\ti\theta\, d\ti\psi^2 +d\ti\theta^2 +\sin^2 \ti\theta d\Omega_3^2
\right \}
\ ,
\label{eq:global}
\end{align}
since the symmetry is manifest in this coordinate system. 

The bosonic part of the string action is given by
\begin{equation}
I={1\over 4\pi\alpha'}\int d\tau\int_{0}^{2\pi\alpha'}d\sigma
\sqrt{-h}h^{\alpha\beta}\partial_{\alpha}X^\mu
\partial_{\beta}X^{\nu}g_{\mu\nu}.
\end{equation}
In addition to the equation of motion for $X^\mu$,
there is a constraint obtained 
by varying the action with respect to the
worldsheet metric $h^{\alpha\beta}$, 
\begin{equation}
-{1\over 2}h_{\alpha\beta}\partial^{\gamma}X^\mu
\partial_{\gamma}X^{\nu}g_{\mu\nu}+\partial_{\alpha}X^\mu
\partial_{\beta}X^{\nu}g_{\mu\nu}=0.
\label{eq:const0}
\end{equation}
In this section we will take the conformal gauge, 
$\sqrt{-h}h^{\alpha\beta}=\eta^{\alpha\beta}$,
in which the above constraint becomes
\begin{align}
\dot{X}^{\mu}\dot{X}^{\nu}g_{\mu\nu}&=-{X}^{\mu}{}'{X}^{\nu}{}'g_{\mu\nu},
\label{eq:const1}\\
\dot{X}^{\mu}{X}^{\nu}{}'g_{\mu\nu}&=0.
\label{eq:const2}
\end{align}

On the background \eqref{eq:global}, 
the momenta corresponding to the translations in 
$t$, $\psi$ and $\ti\psi$, are conserved. We will call them
$E$, $S$ and $J$, respectively,
\begin{align}
E&\equiv P_t\equiv {\delta S\over \delta \dot{t}}={L^2\over 2\pi\alpha'}
\int_{0}^{2\pi\alpha'}d\sigma \cosh^2\rho\, \dot{t},
\label{eq:ESJ1}\\ 
S&\equiv P_\psi\equiv -{\delta S\over \delta \dot{\psi}}={L^2\over 2\pi\alpha'}
\int_{0}^{2\pi\alpha'}d\sigma \sinh^2\rho\, \cos^2\theta \dot{\psi},
\label{eq:ESJ2}\\
J&\equiv P_{\ti\psi}
\equiv -{\delta S\over \delta \dot{\tilde\psi}}={L^2\over 2\pi\alpha'}
\int_{0}^{2\pi\alpha'}d\sigma \cos^2\ti\theta\, \dot{\ti\psi}.
\label{eq:ESJ3} 
\end{align}

Following GKP~\cite{GKP2} (also allowing the motion
along $S^5$~\cite{Tseytlin1, Tseytlin2, Tseytlin3}), 
we take the following ansatz for classical solutions
(with $\theta=\tilde\theta=0$),
\begin{equation}
t=\tau,\quad \psi=\omega \tau,\quad \rho=\rho(\sigma),\quad 
\ti\psi=\tilde\omega\tau.
\label{eq:ansatz}
\end{equation}
Since $t$ and $\psi$ are functions of $\tau$ only, and $\rho$ is a function
of $\sigma$ only, the constraint \eqref{eq:const2} is satisfied. The other 
constraint \eqref{eq:const1} gives the relation, 
\begin{equation}
d\sigma={d\rho\over \sqrt{\cosh^2\rho-\tilde\omega^2-\omega^2\sinh^2\rho}}\, ,
\label{eq:sigmarho}
\end{equation}
which implicitly determines $\rho$ as a function of $\sigma$. 
The string is folded and stretched: the points   
$\sigma=0$ and $\sigma=\pi{\alpha}'$ on the worldsheet
are at the center of the string, $\rho=0$; the points 
$\sigma=\pi{\alpha}'/2$ and $\sigma=3 \pi{\alpha}'/2$
are at the end, $\rho=\rho_0$.

From \eqref{eq:ESJ1}, \eqref{eq:ESJ2} and \eqref{eq:ESJ3}, 
one can find the relations 
among $E$, $S$, $J$. For example, as described in \cite{GKP2}
(for $J=0$), for small $S$ one has $E^2\sim S$;  
for large $S$ one has $E-S\sim \ln S$. By identifying the energy
with the scaling dimension, $\Delta\sim E$, strong coupling 
results of gauge theory have been obtained~\cite{GKP2}.

\subsection{Euclidean signature}
\label{EuclideanString}
We would like to relate the GKP's string solution~\cite{GKP2}
with the GKPW prescription~\cite{WittenAdS, GKP} 
based on supergravity. Since the latter is 
defined on Euclidean AdS, we replace $t\to it_{\rm E}$ in \eqref{eq:global}. 
To study strings (or particles) on this background,
we will take the worldsheet (or worldline) time imaginary also, 
$\tau\to i\tau_{\rm E}$. In this Euclidean calculation, the angular 
momenta $J$ and $S$ should be kept real, since these are 
quantum numbers that specify 
the representation of the symmetry group, and have direct meanings
in gauge theory. Accordingly, as we see from \eqref{eq:ESJ2} and 
\eqref{eq:ESJ3}, the angular
variables $\psi$ and $\ti\psi$ should be taken imaginary\footnote{%
We regard the momentum representation to be fundamental for some
variables such as $\psi$ and $\ti\psi$ here. Thus, we do not particularly
pursue physical meanings for the imaginary values of $\psi$ and $\ti\psi$.}, 
since $\tau$ is now imaginary. This is a straightforward extension of
the prescription of Dobashi, Shimada and Yoneya~\cite{DSY}, in which
one angle was taken to be imaginary, to the case of two angles.
The Euclidean solution takes the form, 
\begin{equation}
t_{\rm E}=\tau_{\rm E},\quad \psi=i\omega \tau_{\rm E},\quad 
\rho=\rho(\sigma),\quad \ti\psi=i\tilde\omega \tau_{\rm E},
\label{eq:ansatzE}
\end{equation}
with $\omega$ and $\tilde\omega$ being real; these are the same as 
the angular velocities for the Lorentzian solution.

By solving the constraint \eqref{eq:const1}, which now becomes
\begin{equation}
\partial_{\tau_{\rm E}}X^{\mu} \partial_{\tau_{\rm E}}X^{\nu}g_{\mu\nu}
={X}^{\mu}{}'{X}^{\nu}{}'g_{\mu\nu},
\label{eq:const1E}
\end{equation}
we obtain the relation between $\rho$ and $\sigma$. It is of 
the same form as \eqref{eq:sigmarho}.

The Euclidean version of $E$, which we will call $H$, is
\begin{equation}
H={L^2\over 2\pi\alpha'}\int_{0}^{2\pi\alpha'}d\sigma\cosh^2\rho 
{dt_{\rm E}\over d\tau_{\rm E}}
={2L^2\over \pi\alpha'}\int_{0}^{\rho_0}d\rho 
{\cosh^2\rho\over \sqrt{\cosh^2\rho-\tilde\omega^2-\omega^2\sinh^2\rho}}~,
\label{eq:HE}
\end{equation}
and the angular momentum along $\psi$ is
\begin{equation}
S=-i{L^2\over 2\pi\alpha'}\int_{0}^{2\pi\alpha'}d\sigma\sinh^2\rho 
{d\psi\over d\tau_{\rm E}}
={2L^2\over \pi\alpha'}\omega \int_{0}^{\rho_0}d\rho 
{\sinh^2\rho\over \sqrt{\cosh^2\rho-\tilde\omega^2-\omega^2\sinh^2\rho}}~.
\label{eq:SE}
\end{equation}

\subsubsection*{Poincar\'{e} coordinates}
We now rewrite the above solution in the Poincar\'{e} coordinates,
so that the correspondence with gauge theory becomes clear. 
For clarity, we present the coordinate
transformations using the embedding coordinates explicitly. 
We will write the formulas for general $p$ in this subsection. 

The Euclidean AdS$_{p+2}$ is represented as a hyperboloid,
\begin{equation}
-X_{p+2}^{2}+X_{0}^{2}+\sum_{a=1}^{p+1}X_{a}^{2}=-L^2,
\end{equation}
in the $p+3$ dimensional flat space, $ds^2=-dX_{p+2}^{2}+dX_{0}^{2}+\sum_{a=1}^{p+1}dX_{a}^{2}$. The hyperboloid is parametrized by the global coordinates as
\begin{align}
X_{p+2}&=L\cosh\rho \cosh t_{\rm E},\nonumber\\
X_{0}&=L\cosh\rho \sinh t_{\rm E},\nonumber\\
X_{a}&=L\sinh\rho \Omega_{a},
\label{eq:embed1}
\end{align}
where $\sum_{a=1}^{p+1}\Omega_{a}^{2}=1$. 

The solution \eqref{eq:ansatzE} effectively
has angular variable $\psi$ imaginary. If we choose $\psi$ to 
parametrize the rotation in the $X_{1}$-$X_{2}$ plane as follows,
\begin{equation}
X_{1}=L\sinh\rho \cos\theta \cos\psi,\quad 
X_{2}=L\sinh\rho \cos\theta \sin\psi,
\end{equation}
the coordinate $X_{2}$ effectively becomes imaginary if $\psi$ is 
imaginary. In the following, $X_{2}$, and also $x_{2}$ defined below,
are understood to be imaginary\footnote{%
If we define a real variable 
$\hat{x}_{2}$ by $x_{2}=i \hat{x}_{2}$, the expression
$\sum_{i}x_{i}^2$ below means $x_{1}^2-\hat{x}_{2}^2+x_{3}^2+\cdots$.}.

The Poincar\'{e} coordinates are defined as 
\begin{align}
X_{p+2}&={z\over 2}\left(1+ {L^2+\sum_{i}x_i^2+t_{\rm P}^2 \over z^2}\right),
\nonumber\\
X_{0}&=L {t_{\rm P}\over z},\qquad X_{i}=L {x_{i}\over z},\nonumber\\
X_{p+1}&={z\over 2}\left(1+ {-L^2+\sum_{i}x_i^2+t_{\rm P}^2 \over z^2}\right).
\label{eq:embed2}
\end{align}
Equating \eqref{eq:embed1} and \eqref{eq:embed2}, we obtain
\begin{align}
t_{\rm P}&=\tilde\ell \tanh \tau_{\rm E},
\label{eq:tE}\\
z&={\tilde\ell\over \cosh\rho \cosh\tau_{\rm E}},
\label{eq:z}\\
\sum_{i}x_{i}^2&=z^2\sinh^2\rho={\tilde\ell^2\tanh^2\rho\over \cosh^2\tau_{\rm E}},
\label{eq:xi}
\end{align}
where $\tilde\ell$ is an arbitrary constant, and we have used
$t_E=\tau_E$ from \eqref{eq:const1E}.

From \eqref{eq:xi}, we see that the spatial extent of the string
reduces to a point in terms of the coordinate $x_{i}$, as the string approaches the boundary $z\to 0$ (or $\tau\to\pm\infty$), in the following sense:
the string occupies $0\le \rho\le \rho_0$ (with $\rho_0$ being
a finite constant), and the center of the string, $\rho=0$, is at $x_{i}=0$; 
eq.\ \eqref{eq:xi} shows that
the value of $\sum_{i} x_{i}^2$ corresponding to the endpoint of the 
string, $\rho=\rho_0$, goes to zero\footnote{%
If $x_{i}$ were all real, this would really mean that the string 
reduces to a single point 
at the boundary. In our case where $x_{2}=i\hat{x}_{2}$ is imaginary, 
string is really stretched in the ``lightlike'' direction in the 
$x_{1}$-$\hat{x}_{2}$ plane. 
However, as mentioned in a previous footnote, we do not attach 
a particular physical meaning to imaginary values of the coordinates. 
Thus, if the invariant distance $\sum_{i}x_{i}^2$ is zero, we interpret
it as a single point.} 
as $z\to 0$ (or $|\tau|\to \infty$).
This fact is in comfort with the fact that we represent this state
as a local operator in gauge theory. 

\subsubsection*{Gauge-theory correlator}
By substituting $\tau_{\rm E}=\pm\infty$ in \eqref{eq:tE}, we see that
the coordinate distance between the two points, $t_{\rm i}$ and $t_{\rm f}$
(both with $x_i=0$), on the boundary is given by 
\begin{equation}
|t_{\rm f}-t_{\rm i}|=2\tilde\ell.
\end{equation}
In fact, \eqref{eq:xi} shows that the center of mass of the string follows 
the same trajectory as the trajectory of a particle which has an angular 
momentum along $S^{8-p}$ (and not along AdS$_{p+2}$). The latter was 
first studied for $p=3$ by Dobashi, Shimada and Yoneya~\cite{DSY}, and 
then generalized to $p\neq 3$ by Asano, Sekino and Yoneya~\cite{ASY}.
It will be also described in Sec.~4 of this paper. 

The Euclidean amplitude for a string which propagates from a point 
on the boundary to another point
on the boundary is interpreted as the two-point function of gauge theory.
In the classical approximation, the bulk amplitude is
\begin{equation}
e^{-S_{\rm cl}}=e^{-\int_{-T}^{T} d\tau_{\rm E} H(\tau_{\rm E})},
\end{equation}
where $H(\tau_{\rm E})$ is the Hamiltonian at Euclidean worldsheet time 
$\tau_{\rm E}$. This is just the Euclidean version of the global energy
given in \eqref{eq:HE}, since the worldsheet time is equal to the global time,
in our classical solution \eqref{eq:ansatzE}. 

We introduce a cutoff $T$ for
the worldsheet time, $-T\le \tau \le T$. We also introduce a cutoff for the
radial coordinate, $z\ge 1/\Lambda$. This IR cutoff in the bulk is interpreted
as a UV cutoff in gauge theory~\cite{SusskindWitten}. 
The relation between the two cutoffs can be
read off from \eqref{eq:tE} and \eqref{eq:z} in the $|T|\to \infty$ limit
as 
\begin{equation}
2\tilde\ell \Lambda =  e^{T}.
\end{equation}

For the solution \eqref{eq:ansatzE}, the Hamiltonian is independent of the
worldsheet time. Therefore, it can be taken out of the integral, and 
amplitude can be written as
\begin{equation}
e^{-\int_{-T}^{T} d\tau_{\rm E} H(\tau_{\rm E})}=e^{-2H T}=
{1\over \left(\Lambda|t_{\rm f}-t_{\rm i}|\right)^{2H}}.
\end{equation}
The worldsheet Hamiltonian gives the scaling dimension, and 
we have recovered the result of GKP. 

This formalism should be applicable to the $p\neq 3$ case without 
conformal symmetry.  
In that case, the worldsheet Hamiltonian will not be 
time independent in general, thus the above integral 
has to be evaluated explicitly. 
In addition, finding the string solution is more 
challenging than $p=3$, since the AdS isometry is not the
symmetry of the string due to the position-dependent Weyl
factor. One approach would be to study the limit of short 
string, by using the approximate form of the geometry near
the center of the string.
This subject is under study~\cite{WIP},
and will be reported elsewhere.

\section{Rotating particles (weak gauge coupling)}
\label{WeakCoupling}
Let us now consider weak gauge coupling. In this paper,
we will concentrate on the case where the gauge coupling is strictly zero. 
On the string theory side, this corresponds to the case where
string tension is zero. 
If the spatial direction of the worldsheet is discretized into bits, 
the interactions among the bits can be ignored in this limit. 
The previous paper~\cite{Sekino} considered the state which has
angular momentum only along $S^{8-p}$, but here we will extend
the analysis to those which have angular momentum also 
along AdS$_{p+2}$. As in~\cite{Sekino}, we assume that one bit 
carries\footnote{%
Here, we are considering the orbital angular momentum, since a
bit (particle) is pointlike. This is in contrast to 
the stretched string spinning around its center of mass,
studied in the previous section.}
a single unit of angular momentum along $S^{8-p}$. 

We first study the $p=3$ case using the Lorentzian formulation, by identifying
the AdS energy in terms of the global time with the scaling dimension. We then
study the case of general $p$ using the Euclidean formulation 
which does not rely on conformal symmetry. 

\subsection{$p=3$}
A bit is a massless particle in ten dimensional spacetime~\cite{Sekino},
namely, AdS$_{5}\times S^5$ in this case. 
We will fix the angular momentum in the $S^5$ direction. Then,
we can perform Kaluza-Klein reduction, and treat
a bit as a massive particle on AdS$_{5}$. 
More precisely, by considering the Routh function (in which Legendre 
transformation is performed for an angle in the $S^5$ direction) 
in ten dimensions, we obtain the action of massive particle on
AdS$_5$ (see Sec.~2.2 of the paper~\cite{ASY}). 
The mass is given by $m=J/L$, 
where $J$ is an integer, and $L$ is the radius of $S^5$, 
which is equal to the radius of AdS$_5$. One  
bit has $J=1$~\cite{Sekino}, but we will keep $m$ in the formulas below
for the time being.

The action for a single bit is given by 
\bea
I_{\rm bit}[X^{\mu}] = \frac{1}{2} \int d\tau \left[ \frac{1}{\eta(\tau)} g_{\rho\sigma}(X^{\mu}(\tau)) \dot{X}^{\rho}(\tau) \dot{X}^{\sigma}(\tau) - \eta(\tau) m^2  \right], 
\ena 
and the corresponding equations of motion are
\bea
g_{\rho\sigma}\dot{X}^{\rho}\dot{X}^{\sigma}&=& - \eta^2 m^2, 
\label{eq:p3constraint}\\
\dot{X}^{\rho}\pd_{\rho} \dot{X}^{\mu} &=& - \Gamma^{\mu}{}_{\rho\sigma}  \dot{X}^{\rho}\dot{X}^{\sigma}.
\ena
For AdS$_{5}$ in the global coordinates \eqref{eq:global}, 
the single-bit action becomes
\bea
I_{\rm bit}[X^{\mu}] = \frac{1}{2} \int d\tau \left[ \frac{L^2}{\eta}\left( -\cosh ^2\rho ~ \dot{t}^2 + \dot{\rho}^2 + \sinh^2 \rho ~ \dot{\psi}^2 + ({\rm remaining ~ angular ~ part})  \right) -\eta m^2 \right]. \notag 
\ena
The following canonical momenta are conserved,
corresponding to the isometry of the background,
\bea
E\equiv - p_{t} = \frac{L^2}{\eta} \cosh^2 \rho ~ \dot{t},  \\
S \equiv p_{\psi} =  \frac{L^2}{\eta} \sinh^2 \rho ~ \dot{\psi}.
\ena

Let us take $\rho={\rm const.}$ as an ansatz for a classical solution.
Setting $\dot{\rho}=0$ in the constraint \eqref{eq:p3constraint},
we find
\bea
- \frac{E^2}{\cosh^2 \rho} + \frac{S^2}{\sinh^2 \rho} 
= -m^2 L^2 ~.\label{A}
\ena
In addition, in order to have $\ddot{\rho}=0$,  
the equation of motion for $\rho$,
\bea
\ddot{\rho}& =& \cosh \rho \sinh \rho \left[ -\dot{t}^2 + \dot{\psi}^2 \right] \hspace{3.35cm} \notag \\
&=& \cosh \rho \sinh \rho\, \frac{\eta^2}{L^4} \left[ -\frac{E^2}{\cosh^4 \rho} + \frac{S^2}{\sinh^4 \rho} \right] ~ , 
\ena
indicates 
\bea
-\frac{E^2}{\cosh^4 \rho} + \frac{S^2}{\sinh^4 \rho} = 0 ~ .\label{B}
\ena
Eqs.\ (\ref{A}) and (\ref{B}) lead to\footnote{%
The energy takes only positive values $E\ge 0$, but the 
angular momentum $S$ could be positive or negative.} 
\bea
E^2 = {m^2 L^2} \cosh^4 \rho ~~, ~~ S^2 = {m^2 L^2} \sinh^4 \rho \notag \\
\Longleftrightarrow ~~ 
E = {m L} \cosh^2 \rho ~~, ~~ |S| = {m L} \sinh^2 \rho~.
\ena
Thus,
\bea
E= |S| + {m L}
\ena
One bit has a single unit of angular momentum on $S^5$, and $m=1/L$, 
so its energy is
\bea
E_{\rm bit}=|S|+1~.
\ena

Now consider a collection of $n$ non-interacting bits. Its 
total energy $\displaystyle E=\sum_{i=1}^{n}E_{i}$ can be written as
\bea
E=\sum_{i=1}^{n}|S_{i}| + n,
\label{eq:nbitsE1}
\ena
where $|S_{i}|$ is the magnitude of the angular momentum along
AdS$_{5}$ (which is integer in our convention) carried by the
$i$-th bit. The $n$ bits can have different directions
of angular momenta. 
One bit has a single unit of angular momentum
along $S^5$, and contributes 1 to the energy; 
summing this over the $n$ bits,
we obtain $n$ in the last term of \eqref{eq:nbitsE1}. 
In the special case where all the bits have angular momenta in 
the same direction both for AdS$_{5}$ and $S^5$,
which corresponds to an operator\footnote{%
Here we only mean that the number of the 
$Z$ field in the trace is $J$, and the number of 
the covariant derivative $D$ is $S$. At strictly zero gauge 
coupling considered in this paper, it would be difficult 
to distinguish different operators within this class.}
such as
${\rm Tr} (Z^{J-k}D^{S}Z^{k})$, the total energy is 
\bea
E=|S|+ J, 
\label{eq:nbitsE2}
\ena
where $|S|$ and $J$ are the magnitudes of the total angular 
momenta along AdS$_{5}$ and $S^5$, respectively.

The above results for the non-interacting bits correctly reproduce
the free-field results in gauge theory, if we identify the global
energy $E$ with the scaling dimension $\Delta$ of the corresponding 
operator. A single bit is assumed to correspond to a
single scalar field inside the trace, which contributes 1 to
the scaling dimension in the free theory in (3+1) dimensions.
The $i$-th bit with angular momentum $S_{i}$ along 
AdS$_{5}$ can be realized by applying $|S_{i}|$ gauge covariant
derivatives on the $i$-th scalar field. In the free theory,
covariant derivative is just a partial derivative, which 
contributes 1 to the scaling dimension.  
Thus, \eqref{eq:nbitsE1} (or \eqref{eq:nbitsE2} for a special case)
gives the correct free-field result under the 
identification $\Delta=E$. 

\subsection{General $p$}
Let us now consider the case of general $p$. We assume $p\ge 2$,
so that we have two spatial directions in gauge theory
in which the string (or particle) can rotate.

The background geometry is conformal to the AdS$_{p+2} \times S^{8-p}$ 
spacetime\footnote{%
Up to now we have put tilde on the angles in $S^{8-p}$,
but in this subsection, we will omit tilde and denote them
$\theta$, $\psi$, to simplify the notations.},
\bea
ds^2= L^2 r^{-\frac{3-p}{2}} \left[ \left( \frac{2}{5-p} \right)^2 \frac{1}{z^2}\left\{ -dt^2 + dz^2 + dx_{a}^2 \right\} + (d\theta^2 + \cos^2 \theta d\psi^2 +\sin^2 \theta d\Omega_{6-p}^2  )  \right]~ ,
\label{eq:generalpmetric}
\ena
where $a=1,\ldots, p$. 
For general $p$, there is no conformal symmetry (or AdS isometry), 
and we cannot use the identification $\Delta=E$, so we follow the 
DSY prescription~\cite{DSY, ASY, AsanoSekino}. As explained in Section~3,
gauge-theory correlators are obtained by calculating  
the transition amplitude from the path integral in the multiply 
Wick rotated background:
\bea
\langle t_{f}^{E}, J, \Theta_{f}|t_{i}^{E}, J, \Theta_{i}\rangle = \langle it_{f}, J, \Theta_{f}|it_{i}, J, \Theta_{i}\rangle
= \int \MC{D} X  e^{-(I+J\psi_{f}-J\psi_{i}+S\Theta_{f}-S\Theta_{i})} ~. 
\ena
The role of the term $J\psi_{f}-J\psi_{i}$ (or $S\Theta_{f}-S\Theta_{i}$)
on the exponent on the
right hand side is to fix the
angular momentum in the $\psi$ (or $\Theta$) 
direction to $J$. This term can be equivalently
represented by using the Routh function defined by $I+\int d\tau
 (J \dot{\psi}+S\dot{\Theta})$ 
instead of the action $I$. See Sec.~2.2 of \cite{ASY}.

As in the last subsection, we will study the case of zero gauge coupling
by ignoring interactions among the bits. The action for a single bit
on the background \eqref{eq:generalpmetric} is
\bea
I = \int_{-T}^{T} d\tau  ~ \frac{\ti{\alpha}}{2}L^2 \left[ \left( \frac{2}{5-p}\right)^2 \frac{1}{z^2} \left\{ \dot{t}^2 + \dot{z}^2 +\dot{R}^2 - R^2 \dot{\Theta}^2 + \cdots \right\} + (\dot{\theta}^2 - \cos^2 \theta \dot{\psi}^2 + \cdots) \right].  
\label{eq:generalpI}
\ena
The factor $\ti{\alpha}$ is a constant
to be determined later. 
The coordinates $R$ and $\Theta$ are the radius and angle 
defined from two of the coordinates $x_{a}$. 
 
Here we have absorbed the Weyl factor in the metric \eqref{eq:generalpmetric}
by a choice of the einbein (which is an overall factor in the action, since
we are considering massless particles in 10D)~\cite{ASY, AsanoSekino}. 
As a result, the action \eqref{eq:generalpI} is formally the same
as the one in AdS$_{p+2}\times S^{8-p}$. In string theory this amounts 
to taking the worldsheet metric so that
$\sqrt{h}h^{\tau\tau}=\bar{r}^{{\frac{3-p}{2}}}$ (with $h^{\tau\sigma}=0$), 
to absorb the overall
factor $\bar{r}^{-{\frac{3-p}{2}}}$ in the metric. Then, 
$\sqrt{h}h^{\sigma\sigma}=1/(\sqrt{h}h^{\tau\tau})=\bar{r}^{-{\frac{3-p}{2}}}$,
and the coefficient of $(\partial_\sigma x^{i})^2$ in the worldsheet action 
gets a time dependent factor 
$\bar{r}^{-(3-p)}$~\cite{ASY, AsanoSekino}\footnote{%
If one takes the conformal gauge $\sqrt{h}h^{\alpha\beta}
=\delta^{\alpha\beta}$,
both of $(\partial_\tau x^{i})^2$ and $(\partial_\sigma x^{i})^2$ have time
dependent coefficients, but the final result of the calculation is of course
independent of the gauge choice~\cite{ASY, AsanoSekino}.}.
Thus, the effect of the
Weyl factor appears for excited states of the string (for which
$\partial_\sigma x^{i}\neq 0$). In the bit string picture, coefficients 
of the interaction terms among the bits will have time dependence 
for $p\neq 3$. 

The equations of motion are
\bea
&&\frac{d}{d\tau} \left( \frac{\dot{t}}{z^2} \right)=0, ~~ \frac{d}{d\tau} \left( \frac{\dot{z}}{z^2} \right)=- \frac{1}{z^3}\{ \dot{t}^2 + \dot{z}^2 + \dot{R}^2 -R^2 \dot{\Theta}^2 + \cdots \}, ~~ \frac{d}{d\tau} \left( \frac{\dot{R}}{z^2} \right)=- \frac{R}{z^2} \dot{\Theta}^2, ~~   \notag \\
&&\frac{d}{d\tau} \left( \frac{R^2}{z^2} \dot{ \Theta} \right)=0, ~~ \frac{d}{d\tau} \dot{\theta} = -\sin \theta \cos \theta \dot{\psi}^2, ~~ \frac{d}{d\tau} \dot{\psi} =0, ~~
\ena
and the constraint is 
\bea
\left( \frac{2}{5-p}\right)^2 \frac{1}{z^2} \left\{ \dot{t}^2 + \dot{z}^2 +\dot{R}^2 - R^2 \dot{\Theta}^2 + \cdots \right\} + (\dot{\theta}^2 - \cos^2 \theta \dot{\psi}^2 + \cdots) =0~.
\ena

\subsubsection*{The $S=0$ case}
When $S=0$, the solution has been obtained by Asano, Sekino and 
Yoneya~\cite{ASY, AsanoSekino};
\bea
&&t= \ti{\ell} \tanh \tau ,  ~~ z= \frac{\ti{\ell}}{\cosh \tau} , ~~ R=\Theta=0, \notag \\
&&\theta = 0 , ~~ \psi= \frac{2}{5-p}\tau~,
\label{eq:ASYsolution}
\ena 
where $\ti{\ell}$ parametrizes the separation between $t_{f}$ and $t_{i}$, 
that is,
\bea
|t_{f}-t_{i}|= 2 \ti{\ell} ~.
\label{eq:tfti}
\ena
The solution \eqref{eq:ASYsolution} is represented as a half circle
$t^2+z^2=\ti{\ell}^2$ in the coordinate space $(t, z)$. 

As in the last section,
we introduce the IR cutoff at $z=\frac{1}{\Lambda}$, which is related
to the worldsheet cutoff $T$ as 
\bea
2\ti{\ell} \Lambda =  e^{T}.
\label{eq:llambda}
\ena
The transition amplitude at the zero-loop level on the worldsheet 
becomes
\bea
\langle t_{f}, J, 0|t_{i}, J, 0\rangle &\simeq &e^{-(I+J(\psi_{f}-\psi_{i}))} 
= e^{- \frac{4}{5-p}JT}
\notag \\
& = &
\frac{1}{\Lambda^{\frac{4}{5-p}J}} \frac{1}{|t_{f}-t_{i}|^{\frac{4}{5-p}J}}
 ~~~~ {\rm ~ (at ~ zero~loop)}
\label{eq:zeroloop}
\ena
where we have used \eqref{eq:tfti} and \eqref{eq:llambda} in the last line
to rewrite the amplitude in terms of the variables in gauge theory. We have
the factor $J$ on the exponent, because there are $J$ non-interaction bits,
each giving the contribution described above. 
The expression \eqref{eq:zeroloop} could be regarded as the 
leading part in the large $J$ limit of the correlator of ${\rm Tr}(Z^J)$
at strong gauge coupling~\cite{ASY, AsanoSekino}. 

Let us consider the one-loop contribution on the worldsheet, following 
\cite{Sekino}.  
The action of a single bit at the quadratic level of the bosonic 
fluctuations around the classical trajectory~\eqref{eq:ASYsolution} 
is~\cite{ASY, AsanoSekino}
\bea
I^{(2)}= \frac{\ti{\alpha}}{2} \int_{-T}^{T} d\tau \left\{ \dot{x}_{a}^2 + m_{x}^2 x_{a}^2 + \dot{y}^2_{i} + m_{y}^2 y_{i}^2   \right\}\, . \label{secondorder}
\ena
We have $(p+1)$ fields $x_{a}$ with mass $m_{x}=1$ which comes from 
fluctuations along AdS$_{p+2}$, and $(7-p)$ fields $y_{i}$ with mass
$m_{y}= \frac{2}{5-p}$ which come from fluctuations along $S^{8-p}$. 
There are also 8 fermionic fluctuations 
with mass $m_{f}={(7-p)\over 2(5-p)}$.
For the quadratic action of the fermionic fluctuations, 
see~\cite{AsanoSekino}.

In \cite{Sekino}, the modification to the correlator \eqref{eq:zeroloop} 
for ${\rm Tr}(Z^J)$ due to the 
one-loop contribution on the worldsheet has been obtained by an operator
method,  by including 
the zero-point energies of the bosonic and fermionic fluctuations
(which are harmonic oscillators) for each bit. 
For $p=3$, the contributions from the
bosonic and fermionic fluctuations cancel each other~\cite{AsanoSekino}, 
since there is effectively worldsheet supersymmetry. 

Here we will derive the same result as~\cite{Sekino}
from the Euclidean path integral of harmonic oscillators 
with the boundary condition $x(T)=x(-T)=0$.
Since the complete set of basis may be given by
\bea
\left\{ \cos \frac{\pi}{2T}(2\bar{k}+1)\tau, ~~ \sin \frac{\pi}{2T}(2\ti{k})\tau \right\}_{\bar{k},\ti{k}}
\ena
the eigenvalues of the operator $\left( -\frac{d^2}{d\tau^2} + m^2 \right)$ are given by 
\bea
\lambda_{k}= \frac{\pi^2}{4T^2} k^2 + m^2  ~~ (k \in \MB{Z}_{\geq0})~.
\ena
Therefore, the path integral is 
\bea
Z^{(2)}=\int \MC{D}x ~ e^{-\frac{1}{2}\int_{-T}^{T}d\tau \left\{ \dot{x}^2 + m^2 x^2\right\} } \hspace{4.7cm} \notag \\
= \prod_{\bar{k},\ti{k}} \int dA \int dB e^{-A^2 \left( \frac{\pi^2}{4T^2} (2\bar{k}+1)^2 + m^2 \right)T} e^{-B^2 \left( \frac{\pi^2}{4T^2} (2\ti{k})^2 + m^2 \right)T}\notag \\
= \prod_{k} \sqrt{\frac{\pi}{\lambda_{k}T}} \hspace{7.5cm}
\ena
that is,
\bea
\log Z^{(2)} = -\frac{1}{2} \sum_{k} \log \lambda_{k} + 
({\rm divergent ~ part})~.
\ena
Here, we apply the zeta function regularization method. Redefine $Z^{(2)}$ by
\bea
\log Z^{(2)} \equiv \frac{1}{2} \left. \frac{d \zeta(s)}{ds}\right|_{s=0}
\ena
where the generalized zeta function is given by
\bea
\zeta(s)\equiv \sum_{k} \lambda_{k}^{-s}~.
\ena
Now make $k$ continuous  by considering large $T$, i.e., 
\bea
\zeta(s) = \sum_{k} \left( \frac{\pi^2}{4T^2}k^2 +m^2 \right)^{-s}
 = \frac{2T}{\pi}\sum_{K} \left( K^2 +m^2 \right)^{-s} \Delta K \notag\\
 \to \frac{2T}{\pi} \int_{0}^{\infty} dK \left( K^2 +m^2 \right)^{-s}
\ena
where we changed the variable $K\equiv \frac{\pi}{2T}k$ and $\Delta K = \frac{\pi}{2T}$. Then, 
\bea
\zeta(s) = \frac{2T}{\pi} \int_{0}^{\infty} dK \left( K^2 +m^2 \right)^{-s} = \frac{T}{\sqrt{\pi}} m^{1-2s} \frac{\Gamma(s-\frac{1}{2})}{\Gamma(s)} \\
\simeq -2mT s + 4mT (-1+\log 2 + \log m) s^2 + \cdots \hspace{0.6cm}
\ena
Eventually,
\bea
Z^{(2)}= e^{\frac{1}{2} \left. \frac{d \zeta(s)}{ds}\right|_{s=0}}
 = e^{-mT}~.
\label{eq:Z2}
\ena
Note that this is independent of the overall factor of the harmonic oscillator action. Therefore, the one-loop contribution of the bosonic part for $J$ bits is
\bea
Z^{(2)}_{J\, {\rm bit(boson)}}= e^{-J ((p+1)m_{x}+(7-p)m_{y})T}=e^{-\frac{-p^2 + 2p +19}{5-p} JT}~.
\ena
 
Similarly, the one-loop contribution of the fermionic part is
\bea
Z^{(2)}_{J\, {\rm bit(fermion)}}=e^{J 8 m_{f}T}= e^{\frac{28-4p}{(5-p)}JT} ~.
\ena 
Then
\bea
Z^{(2)}_{J\, {\rm bit}}=Z^{(2)}_{J\, {\rm bit(boson)}}
Z^{(2)}_{J\, {\rm bit(fermion)}}
=e^{\frac{(p-3)^2}{5-p}JT} ~.
\ena

Combining the zero- and one-loop 
contributions on the worldsheet, the amplitude becomes
\bea
\langle t_{f}, J, 0|t_{i}, J, 0\rangle &\simeq& e^{-(I+J(\psi_{f}-\psi_{i}))}Z^{(2)}_{J\, {\rm bit}} \hspace{3.5cm} \notag \\
& =& e^{- (p-1)JT} \hspace{5.25cm} \notag \\
& =& \frac{1}{\Lambda^{ (p-1)J}} \frac{1}{|t_{f}-t_{i}|^{ (p-1)J}}.
\ena
This reproduces the free-field result for the two-point function 
of ${\rm Tr}(Z^J)$
in the $(p+1)$-dimensional gauge theory~\cite{Sekino}: 
from dimensional analysis,
a scalar field has dimension $(p-1)/2$ in the free theory, 
and the operator consists of $J$ scalar fields.

\subsubsection*{The $S\neq 0$ case}
In this case, the solution is 
\bea
&&t= \ti{\ell} \tanh \tau , ~~~ z= \frac{\sqrt{\ti{\ell}^2-B^2}}{\cosh \tau},  ~~~ R= \frac{B}{\cosh \tau},  ~~~ \Theta= \tau, \notag \\
&&\theta = 0, ~~~ \psi= \frac{2}{5-p}\tau ~.
\label{eq:Sneq0solution}
\ena
Similarly to the solution \eqref{eq:ASYsolution} for $S=0$, this is 
represented 
as a half circle, now given by $t^2+z^2+R^2=\ti{\ell}^2$ 
in the coordinate space $(t, z, R)$.
Since $J$ and $S$ are given by
\bea
&& J= \ti{\alpha} L^2 \dot{\psi}= \ti{\alpha} L^2 \frac{2}{5-p}~, \\
&& S = \ti{\alpha} L^2 \left( \frac{2}{5-p} \right)^2 \frac{R^2}{z^2} \dot{\Theta} =  \ti{\alpha} L^2 \left( \frac{2}{5-p} \right)^2  \frac{B^2}{\ti{\ell}^2 - B^2}~,
\ena
 the constants $\ti{\alpha} $ and $B$ are related to $J, S$ and $\ti{\ell}$ via
\bea
&& \ti{\alpha} = \frac{5-p}{2}\frac{J}{L^2}~, \\
&& B= \sqrt{\frac{S(5-p)}{2J+S(5-p)}} \ti{\ell} ~ .
\ena

We introduce the IR cutoff at $z=\frac{1}{\Lambda}$. The relation 
between $\Lambda$ and $T$ can be read off from \eqref{eq:Sneq0solution}
in the $|T|\to \infty$ limit, and becomes
\bea
2\ti{\ell} \Lambda \sqrt{\frac{2J}{2J + S(5-p)}} = e^{T} ~.
\ena
At the zero-loop level on the worldsheet, 
the transition amplitude is given by
\bea
\langle t_{f}, J, S|t_{i}, J, S\rangle \simeq e^{-(I+J(\psi_{f}-\psi_{i})+S(\Theta_{f}-\Theta_{i}))} \hspace{4.65cm} \notag \\
 = e^{- (\frac{4}{5-p}J+2S)T} \hspace{6.45cm} \notag \\
 = \left(\frac{2J+S(5-p)}{2J}\right)^{\frac{2}{5-p}J+S}  \frac{1}{\Lambda^{\frac{4}{5-p}J+2S}} \frac{1}{|t_{f}-t_{i}|^{\frac{4}{5-p}J+2S}}~.  \\ 
~~~~ {\rm ~ (at ~ zero~loop)} \notag
\ena

The amplitude including the one-loop contribution can be obtained 
by expanding the coordinates of the particle to the quadratic order 
in fluctuations around the classical solution \eqref{eq:Sneq0solution}
by doing an analysis similar to the one for $S=0$~\cite{AsanoSekino}.
As described in Appendix~\ref{quadratic}, we find the following 
spectrum of the bosonic fluctuations: there are 
$(p-1)$ fields with mass 1 (from the fluctuations along $S^{p}$ within
AdS$_{p+2}$),
one field with mass 2 (from the one along the radial
 direction $\rho$ in AdS$_{p+2}$), 
and $(7-p)$ fields with mass ${2\over 5-p}$ (from the ones along $S^{8-p}$).
We have eight fermionic fluctuations with mass ${7-p\over 2(5-p)}$. 
This spectrum is in contrast to the $S=0$ case\footnote{%
The reason we have one less number (seven) of bosonic fluctuations than
in the $S=0$ case (eight) is that we are fixing one more angular momentum
here, introducing one more constraint among fluctuations.},
as described below
\eqref{secondorder}. Here, we have two less fields with mass 1
than in the $S=0$ case, but have one extra field with mass 2. These  
compensate with each other, making 
the zero-point energy (the one-loop contribution) 
for $S\neq 0$ equal to the one for $S=0$.

By including the one-loop contribution, the amplitude becomes
\bea
\langle t_{f}, J, S|t_{i}, J, S\rangle \simeq e^{-(I+J(\psi_{f}-\psi_{i})+S(\Theta_{f}-\Theta_{i}))}Z^{(2)}_{J-bit}  \hspace{4.05cm} \notag \\
 = e^{- ((p-1)J+2S)T} \hspace{6.65cm} \notag \\
 = \left(\frac{2J+S(5-p)}{2J}\right)^{\frac{p-1}{2}J+S}  \frac{1}{\Lambda^{(p-1)J+2S}} \frac{1}{|t_{f}-t_{i}|^{(p-1)J+2S}} ~.  
\ena
This agrees with the free field result of the correlator for the
operator of the form ${\rm Tr} (Z^{J-k}D^{S}Z^{k})$: At zero coupling,
the covariant derivative $D$ is just a partial derivative. Inserting 
one derivative in the operator increases two powers of the coordinate 
distance in the two-point function.

\section{Conclusions}

In this paper, we considered gauge/gravity correspondence
between maximally supersymmetry Yang-Mills theories in 
($p+1$)-dimensions and superstrings on the near horizon 
limit of the D$p$-brane solutions. 
We computed two-point functions of operators with angular momentum 
along the AdS directions from the string worldsheet theory. 

First, we considered the conventional continuum string theory, which should
correspond to the strongly-coupled gauge theory. We considered
the conformally invariant case of $p=3$, and first considered
the folded and spinning string solution 
found by Gubser, Klebanov and Polyakov~\cite{GKP2}. We rotated some
coordinates to imaginary, and obtained a string configuration 
which connects two points on the boundary. 
We explained that gauge-theory correlators can be obtained as
transition amplitudes for the Euclidean string.
This formalism is not based on the
identification of the energy with the global time, and can be applied
to theories without conformal symmetry. We will defer the analysis
of $p\neq 3$ for future study. This is an interesting technical 
challenge, since the isometry of 
AdS$_{p+2}\times S^{8-p}$ is not a symmetry of string action due
to the position dependent Weyl factor.

Then, we considered the limit of zero gauge coupling. 
As in the previous paper~\cite{Sekino}, 
we assumed the string is made of bits, each of 
which has a single unit of angular momentum along $S^{8-p}$. In the
limit of weak gauge coupling, the string tension is small compared 
with the scale of angular momenta. At zero coupling, we computed the 
amplitude by ignoring the interactions among bits. We obtained the 
free-field correlator of gauge theory, extending the result of \cite{Sekino}
to general operators. We regard this result as an indication for the
validity of the approach for weak gauge coupling
initiated in  \cite{Sekino}. 

As mentioned at the end of Introduction, our result is based on
two unproven assumptions. It is important to clarify whether these
assumptions are true. One assumption is
that the near-horizon D$p$-brane background does not receive
$\alpha'$ corrections for general $p$. This does not seem 
very unrealistic, since the geometry for
$p\neq 3$ has a tensor structure similar to $p=3$, which is 
believed to be $\alpha'$ exact~\cite{Banks, Kallosh}. 
Another assumption is that taking into account 
the quantum effect of fluctuations up to the quadratic order 
gives the correct answer (when computing the amplitude for
a string bit). At the moment, we do not have a clear justification
for this. By performing exact quantization
of superparticle in AdS$_{p+2}$ (without the Weyl factor since 
this does not contribute to the single bit action, as explained
in Sec.~\ref{WeakCoupling}), we should be able to clarify this
point. Or we could compute the next order in the expansion.  
Since the contributions from the
bosonic and fermionic fluctuations cancel with each other at
all orders for $p=3$, it might be possible that there is only 
contribution at the quadratic order for $p\neq 3$. 

There are two main directions for future research. 
One direction is towards understanding the weak-coupling limit
of gauge/gravity correspondence~\cite{WIP}. We will be able to incorporate
the interactions among the string bits perturbatively. 
It is a highly important problem whether perturbative expansion 
in string tension agrees with the perturbative expansion in 
gauge theory. If they agree, this can be regarded as a proof of 
gauge/gravity correspondence. 

Another direction is towards understanding of gauge theory without
conformal symmetry at strong gauge coupling. The
string solution along the lines of Sec.~3 
for $p\neq 3$ is under study~\cite{WIP}. 
It is important to study the large $S$ behavior of that
solution. For $p=3$ there is a characteristic large $S$ behavior 
in the scaling dimensions of the form $\log S$. In gauge theory,
this comes from gauge fields propagating in the internal lines.
Large $S$ behavior 
for $p\neq 3$ will give new piece of information for the structure 
of these gauge theories. 

Apart from the above two problems, it would be interesting
to extend the analysis in this paper to more general backgrounds. 
Our formalism of computing the gauge-theory correlator from the 
string amplitude, described in Sec.~3, 
would be applicable to general backgrounds without
conformal symmetry. Also, the approach to weakly coupled gauge theories
based on the string bit picture, described in Sec.~4 and in 
\cite{Sekino}, will be applicable for backgrounds in which 
the two assumptions mentioned above are satisfied.

\section*{Acknowledgments}
This work is supported in part by MEXT KAKENHI Grant Number 21H05187 (YS) and the Waseda University Grant for Special Research Project (No.\ 2021C-570) (SM).

\appendix

\section{Fluctuations around classical solution of a particle}
\label{quadratic}

In this Appendix, we will obtain the massless particle action at 
the quadratic
order in the fluctuations around a classical solution 
on the near-horizon limit of the 
D$p$-brane solution.
We are really interested in the background
with the Euclidean time and imaginary angular coordinates, 
but here we will do the analysis in the usual Lorentzian 
background, since the fluctuations around the former can be 
obtained simply by rotating the time and two angles to imaginary values.

The bosonic part of the 
massless particle action in the 10-dimensional spacetime is
\begin{align}
I_{\rm b}&
= {L^2\over 2} \int d\tau\, \eta^{-1}f^2(r)
\left[ c^2 \left\{
-\cosh^2\rho ~ \dot{t}^2 + \dot{\rho}^2 
+ \sinh^2 \rho ~ \left( \cos^2\theta\, \dot{\psi}^2 +\dot\theta^2
+\sin^2 \theta ~ \dot\Omega_{p-2}^2  \right)\right\}\right. \nonumber\\
&\left.
\qquad \qquad \quad +\cos^2\tilde\theta ~ \dot{\tilde\psi}^2 + \dot{\tilde\theta}^2 
+ \sin^2 \tilde\theta  \dot{\tilde\Omega}_{6-p}^2 \right], 
\label{eq:masslessaction}
\end{align}
where 
\[
c={2\over 5-p}.
\]
As in the main text, the coordinates without the tilde are the ones
for AdS$_{p+2}$, and those with the tilde are the ones for $S^{8-p}$. 
In this appendix, we take the coordinates to be dimensionless,
assuming that they are measured in units of $L=(g_s N)^{1/(7-p)}\ell_s$.
We will use the global
coordinates for AdS$_{p+2}$, since the classical solution
is simple in this coordinate system.
The symbol $\dot\Omega_{p-2}^2$ is a shorthand for the kinetic term along
the $(p-2)$-dimensional sphere. Though we will not need its explicit form
in the present analysis, it is given e.g., by
\[
\dot\Omega_{p-2}^2=\dot\phi_1^2+(\sin^2\phi_1)\dot\phi_2^2+\cdots+
(\sin^2\phi_1\cdots \sin^2\phi_{p-3})\dot\phi_{p-2}^2,
\]
if we use a standard metric on $S^{p-2}$ using angular coordinates 
$\phi_1,\ldots, \phi_{p-2}$. The quantity
$\dot{\tilde\Omega}_{6-p}^2$ is defined similarly.

The equations of motion for $t$, $\psi$, 
$\tilde\psi$ tell us that the global energy $E$, angular momenta $S$ 
and $J$, defined as follows, are conserved: 
\begin{align}
E&\equiv 
P_{t}=-{\partial {\cal L}\over \partial \dot{t}}=L^2 c^2 
\eta^{-1}f^2(r)\cosh^2\rho
\, \dot{t},
\label{eq:Pt}\\
S&\equiv 
P_{\psi}={\partial {\cal L}\over \partial \dot{\psi}}
=L^2c^2\eta^{-1}f^2(r)\sinh^2\rho\cos^2\theta\, \dot{\psi},
\label{eq:Ppsi}\\
J&\equiv 
P_{\tilde\psi}={\partial {\cal L}\over \partial \dot{\tilde\psi}}
=L^2\eta^{-1}f^2(r)\cos^2\tilde{\theta}\, \dot{\tilde{\psi}}.
\label{eq:Ptpsi}
\end{align}
There is the massless constraint obtained by varying the action w.r.t.\
$\eta$, 
\begin{align}
&c^2\left\{-\cosh^2\rho ~ \dot{t}^2 + \dot{\rho}^2 
+ \sinh^2 \rho ~ \left( \cos^2\theta\, \dot{\psi}^2 +\dot\theta^2
+\sin^2 \theta ~ \dot\Omega_{p-2}^2  \right)\right\} \nonumber\\
&\qquad +\left(\cos^2\tilde\theta ~ \dot{\tilde\psi}^2 
+ \dot{\tilde\theta}^2 
+ \sin^2 \tilde\theta  \dot{\tilde\Omega}_{6-p}^2\right)=0. 
\label{eq:constr}
\end{align}
The equation of motion for $\rho$ is
\begin{align}
-\ddot{\rho}+\cosh\rho\sinh\rho \left\{
-\dot{t}^2+\cos^2\theta\, \dot{\psi}^2 +\dot\theta^2
+\sin^2 \theta ~ \dot\Omega_{p-2}^2\right\}=0.
\label{eq:rhoeom}
\end{align}

We will consider the following classical solution,
which represents a particle rotating
in the AdS with the fixed radius $\rho_0$,
and also rotating along $S^{8-p}$:
\begin{align}
&t=\tau,\quad \rho=\rho_0, \quad \psi=\tau, \quad \theta=0, \nonumber\\
&\tilde\psi= c \tau, \quad \tilde\theta=0.
\label{eq:classical}
\end{align}
The worldline metric is chosen as follows,
\begin{equation}
\eta=f^2(\bar{r})=\bar{r}^{-{3-p\over 4}},
\end{equation}
where $\bar{r}$ is a function of $\tau$,
which is given by inserting the classical 
solution \eqref{eq:classical} into the coordinate 
transformations 
(given in the main text) from the global 
coordinates on AdS to the radial 
coordinate $r$. With this choice of $\eta$,
the Weyl factor $f(\bar{r})$ disappears
from the action.
One can see that this solution \eqref{eq:classical}
satisfies the constraint 
\eqref{eq:constr} and the equation of motion \eqref{eq:rhoeom}, 
and that the momenta $E$, $S$, $J$ are conserved.

When we Wick-rotate the spacetime coordinates $t$, $\psi$, $\tilde{\psi}$,
and also the worldline time $\tau$, the classical solution 
\eqref{eq:classical} remains to be a solution. This solution can be expressed
in Poincar\'{e} coordinates by performing the coordinate transformations,
\eqref{eq:embed1}, \eqref{eq:embed2}, and is given by 
\eqref{eq:tE}, \eqref{eq:z}, \eqref{eq:xi} with $\rho=\rho_0$
(or equivalently, as \eqref{eq:Sneq0solution}
with $B=\tilde{\ell}\tanh\rho_0$). This trajectory starts from a point 
on the boundary and returns to another point on the boundary, 
unlike the Lorentzian solution which rotate near the center of AdS. 
Therefore, the singularity at the center for the $p\neq 3$ case
causes no problem.

\subsection{Bosonic fluctuations}
We now expand the fields (coordinates) of the particle in powers of 
$1/L$:
\begin{equation}
x^\mu_{\rm (tot)}=x^\mu_{(0)}+{1\over L}x^\mu+{1\over L^2}x^\mu_{(2)}+\cdots.
\end{equation}
The fields which did not have any subscript above
are really the ``total'' field. 
They will be denoted with the subscript (tot) hereafter.
For notational simplicity, we will omit the 
subscript (1) on the 1st order fluctuations, which will be 
used most often in the following. 
The part $x^{\mu}_{(0)}$ represents the 
classical solution \eqref{eq:classical}.

Before expanding the action, let us first study the constraints
which should be satisfied by the fluctuations. 
The massless constraint \eqref{eq:constr} at the first order 
in $1/L$ becomes
\begin{equation}
c\left\{-(\cosh^2\rho_0)\dot{t}+(\sinh^2\rho_0)\dot{\psi}\right\}+ 
\dot{\tilde{\psi}}=0.
\label{eq:constr1}
\end{equation}

Now, we assume the value of the angular momentum $S$ and $J$ 
are fixed at the zeroth order, and will not change at higher orders,
since these are the quantities that we are interested in.
(On the other hand, we do not fix the value of $E$.)
By expanding $S$ in \eqref{eq:Ppsi} to the first 
order and setting it to zero, we get
\begin{equation}
(\sinh\rho_0)\left\{(2\cosh\rho_0)\rho+(\sinh\rho_0)\dot{\psi}\right\}=0.
\label{eq:S1}
\end{equation} 
Also, by expanding $J$ in \eqref{eq:Ptpsi} to the first order and setting
it to zero, we get
\begin{equation}
\dot{\tilde{\psi}}=0.
\label{eq:J1}
\end{equation}
From \eqref{eq:constr1}, \eqref{eq:S1} and \eqref{eq:J1},
assuming $\rho_0\neq 0$, we obtain
\begin{align}
\dot{t}&=-2(\tanh \rho_0)\rho,
\label{eq:dottrho}\\
\dot{\psi}&=-2\left({1\over \tanh\rho_0}\right)\rho.
\label{eq:dotpsirho}
\end{align}
When $S=0$ (i.e., $\rho_0=0$), we get 
$\dot{t}=\dot{\tilde{\psi}}=0$ from \eqref{eq:constr1} and 
\eqref{eq:J1}. (In that case,
\eqref{eq:S1} is trivially satisfied since $\sinh\rho_0=0$.) 

We will now expand the action as 
\[
I_{\rm b}=L^2 I_{{\rm b}(0)}+L I_{{\rm b}(1)}+I_{{\rm b}(2)}+\cdots.
\]
The action at the first order $I_{(1)}$ vanishes
due to the massless constraint described above.
The terms relevant for the calculation of the second-order action
are as follows\footnote{%
We do not assign any factors of $L$ on $\dot{\Omega}^2_{p-2}$ and 
$\dot{\tilde{\Omega}}^2_{6-p}$,
which are (square of the $\tau$-derivative of) angles whose 
corresponding radius is classically 
zero ($\theta_{(0)}=\tilde{\theta}{}_{(0)}=0$). 
We will assign a factor of $1/L$ to 
the cartesian coordinates $x_a$ and $y_l$ (whose origin corresponds
to $\theta=0$ and $\tilde{\theta}=0$) defined in \eqref{eq:xa} 
and \eqref{eq:yl}. Since $\theta$ and $\tilde{\theta}$
already have a factor of $1/L$, we do not put any factor
of $L$ on $\Omega_{a}$ and $\tilde{\Omega}_{l}$.}, 
\begin{align}
I_{\rm b}={L^2c^2\over 2}\int d\tau
&\left[ -\left\{\cosh^2\rho_0 +(2\cosh\rho_0\sinh\rho_0){\rho\over L}
+(2\cosh^2\rho_0-1){\rho^2\over L^2}+\cdots\right\}
\left(1+{\dot{t}\over L}\right)^2
\right.\nonumber\\
&+{\dot{\rho}^2\over L^2}+\left\{\sinh^2\rho_0 +(2\cosh\rho_0\sinh\rho_0)
{\rho\over L}
+(2\cosh^2\rho_0-1){\rho^2\over L^2}+\cdots\right\}\nonumber\\
&\qquad\qquad\qquad\times
\left\{(1-{\theta^2\over L^2}+\cdots)(1+{\dot{\psi}\over L})^2
+{\dot{\theta}^2\over L^2}+
({\theta^2\over L^2}+\cdots)\dot{\Omega}_{p-2}^2\right\}\nonumber\\
&\left.+{1\over c^2}\left\{(1-{{\tilde{\theta}}^2\over L^2}+\cdots)
(c+{\dot{\tilde{\psi}}\over L})^2
+{\dot{\tilde{\theta}}^2\over L^2}+
({\tilde{\theta}^2\over L^2}+\cdots)
\dot{\tilde{\Omega}}_{6-p}^2\right\}\right].
\label{eq:relevantI}
\end{align}
Taking the quadratic part (the order $L^{0}$ terms), we get
\begin{align}
I_{{\rm b}(2)}={c^2\over 2}\int d\tau
&\left[ -(\cosh^2\rho_0)\dot{t}^2+\dot{\rho}^2
+4(\cosh\rho_0\sinh\rho_0)\rho(-\dot{t}+\dot{\psi})
\right.\nonumber\\
&+(\sinh^2\rho_0)\left\{\dot{\psi}^2+\dot{\theta}^2
+\theta^2\dot{\Omega}_{p-2}^2-\theta^2\right\}\nonumber\\
&\left.+{1\over c^2}\left\{\dot{\tilde{\psi}}^2+\dot{\tilde{\theta}}^2
+\tilde{\theta}^2\dot{\tilde{\Omega}}_{6-p}^2
-c^2\tilde{\theta}^2\right\}\right].
\end{align}
Using the relations \eqref{eq:dottrho}-\eqref{eq:J1} to eliminate
$\dot{t}$, $\dot{\psi}$, $\dot{\tilde{\psi}}$, it becomes
\begin{align}
I_{{\rm b}(2)}={c^2\over 2}\int d\tau
&\left[\dot{\rho}^2-4\rho^2 +(\sinh^2\rho_0)\left(\dot{\theta}^2
+\theta^2\dot{\Omega}_{p-2}^2-\theta^2\right)\right.\nonumber\\
&\left.+{1\over c^2}\left\{\dot{\tilde{\theta}}^2
+\tilde{\theta}^2\dot{\tilde{\Omega}}_{6-p}^2
-c^2\tilde{\theta}^2\right\}\right].
\end{align}
By defining the fields (cartesian coordinates) $x_a$ and $y_l$ as 
\begin{align}
x_a&=c(\sinh\rho_0)\theta \Omega_a \qquad (a=1,\ldots, p-1),
\label{eq:xa}\\
y_l&= \tilde{\theta}\tilde{\Omega}_l\qquad  (l=1,\ldots,7-p),
\label{eq:yl}
\end{align}
where $\Omega_a$ ($\tilde{\Omega}_l$)
denotes a vector with unit length
$\sum_{a=1}^{p-1}\Omega_a^2=1$ ($\sum_{l=1}^{6-p}\tilde{\Omega}_l^2=1$), 
representing a point on $S^{p-2}$ ($S^{6-p}$),
and by making a suitable redefinition of fields by constant
factors, the action becomes
\begin{equation}
I_{{\rm b}(2)}={1\over 2}\int d\tau \left[ \dot{\rho}^2-4\rho^2 
+\sum_{a=1}^{p-1}\left(\dot{x}_{a}^2-x_{a}^2\right)
+\sum_{l=1}^{7-p}\left(\dot{y}_{l}^2-c^2y_{l}^2
\right)\right].
\label{eq:bosonfluc}
\end{equation}

When $S=0$, there were $(p+1)$ fields with mass 1, coming
from the fluctuations in the AdS$_{p+2}$ 
directions~\cite{AsanoSekino}\footnote{%
The result for $S=0$ is obtained as follows. 
We have $\rho_0=0$ and $\dot{t}=\dot{\tilde{\psi}}=0$, 
as explained above. We assign no factors of $L$ on
the fluctuations of the angles on $S^{p}$ whose corresponding
radius is zero classically $\rho_{(0)}=0$, for the same reason as 
mentioned in the previous footnote. The factor in the curly bracket
in the third line of \eqref{eq:relevantI} is now replaced
with $\dot{\Omega}_{p}^{2}$ (without any factor of $L$). 
Then, from the first three lines in \eqref{eq:relevantI}, 
we obtain the part of $I_{{\rm b}(2)}$ containing the fluctuations 
from the AdS$_{p+2}$ directions as follows, 
\[
{c^2\over 2}\int d\tau
\left\{ \dot{\rho}^2-\rho^2 +\rho^2\dot{\Omega}^2_{p}\right\}
={1\over 2}\int d\tau
\sum_{a=1}^{p+1}\left\{\dot{x}^2_{a}-x^2_{a}\right\},
\]
where $x_{a}=c\rho\Omega_{a}$, with $\Omega_a$ ($a=1,\ldots, p+1$)
being a unit vector which parametrizes a point on $S^{p}$.
The part of $I^{(2)}$ containing the fluctuations 
from the $S^{8-p}$ directions
is the same as in the $S\neq 0$ case.}.
By contrast, we have $(p-1)$ fields ($x_a$) with mass 1 
from the angular directions along $S^{p}$ in AdS$_{p+2}$ 
and one field (from the radial direction $\rho$) with 
mass 2 (mass-squared $=4$).
The mass of the fluctuations in the $S^{8-p}$ directions 
is the same as in the $S=0$ case; we have $(7-p)$ fields
with mass $c={2\over 5-p}$.

The fact that there are
one less physical degrees of freedom (seven in total)
of fluctuations than
in the $S=0$ case (eight) is that we are fixing one more angular
momentum $S$ now, introducing an extra constraint \eqref{eq:S1}.
The one-loop quantum contribution to the amplitude 
(the contribution from the zero-point energy of fluctuations) 
is the same as in the $S=0$ case,
since the contribution from the field $\rho$ compensate
for the one less d.o.f.

The quadratic action for the fluctuations around the
tunneling null geodesic with the Euclidean time and
imaginary angular directions is given simply by changing 
the signs of the mass-squared terms in \eqref{eq:bosonfluc}.

\subsection{Fermionic fluctuations}
We will find that the mass of the fermionic fluctuations for 
$S\neq 0$ is the same as
in the $S=0$ case obtained in \cite{AsanoSekino}. 
Here we will summarize only the main points, and refer the reader
to~\cite{AsanoSekino} for more details.

We will consider the Green-Schwarz type action for the
fermionic particle, obtained from the Green-Schwarz 
superstring action at the quadratic order in the 
fermions~\cite{Cvetic}
by ignoring the worldsheet spatial derivative. 
For type IIA theories, the fermionic part of the 
particle action is  
\begin{equation}
I_{\rm f}=-{i\over 2\pi}\int d\tau \Theta^{T}\Gamma_{0} \eta^{-1}
\partial_{\tau} x^{\mu}\Gamma_{\mu}{\cal D}_{\tau}\Theta.
\label{eq:fermionaction}
\end{equation}
We will take the vielbein 
$\eta=f^2(r)$ as in the bosonic case.
$\Theta$ is a 32-component
spinor in 10-dimensional spacetime.
We take a real representation, $\Theta^*=\Theta$. The part
${\cal D}_{\tau}\Theta$ is defined as
\begin{equation}
{\cal D}_{\tau}\Theta=\nabla_{\tau}\Theta
+\Omega \partial_{\tau} x^{\mu}\Gamma_{\mu}\Theta.
\end{equation}
The first term is the covariant derivative defined
in the usual manner,
\begin{equation}
\nabla_{\tau}\Theta=\partial_{\tau}\Theta
+{1\over 4}\partial_{\tau} x^{\mu}\omega_{\mu}{}^{\hat{\nu}\hat{\sigma}}
\Gamma_{\hat{\nu}\hat{\sigma}}, 
\end{equation}
where the indices with the hat ($\hat{\mu}, \hat{\nu},\ldots$) are
those for the local Lorentz frame. 
The ($p+2$)-form field strength (sourced by the D$p$-brane with 
even $p$ for type IIA; for $p=4$, we consider the dual of 6-form,
which is 4-form) enter the action through
\begin{equation}
\Omega=-{1\over 16}e^{\phi}\left(\Gamma_{11}\Gamma^{\mu_1\mu_2}F_{\mu_1\mu_2}
-{1\over 12}\Gamma^{\mu_1\mu_2\mu_3\mu_4}F_{\mu_1\mu_2\mu_3\mu_4}\right).
\end{equation}

We write the vielbein as
\begin{equation}
e_{\mu}{}^{\hat{\nu}}=Lf(\bar{r})\hat{e}_{\mu}{}^{\hat{\nu}},
\label{eq:fvielbein}
\end{equation}
where the vielbein with the tilde $\hat{e}_{\mu}{}^{\hat{\nu}}$ is the
one for AdS$_{p+2}\times S^{8-p}$, represented as 
\begin{align}
\hat{e}^{\hat{t}}&=c\cosh\rho dt,\quad 
\hat{e}^{\hat{\rho}}=cd\rho,\quad
\hat{e}^{\hat{\theta}}=c\sinh\rho d\theta, \quad
\hat{e}^{\hat{\psi}}=c\sinh\rho \cos\theta d\psi, \nonumber\\
\hat{e}^{\hat{a}}&=c\sinh\rho \sin\theta d\Omega_{a},\quad
\hat{e}^{\hat{\tilde{\psi}}}= \cos\tilde{\theta} d\tilde{\psi},\quad
\hat{e}^{\hat{\tilde{\theta}}}=d\tilde{\theta},\quad 
\hat{e}^{\hat{l}}=\sin\tilde{\theta} d\Omega_{l}.
\label{eq:hate}
\end{align}
For the vielbein of the form \eqref{eq:fvielbein},
the spin connection is written as
\begin{equation}
\omega_{\mu}{}^{\hat{a}\hat{b}}
=(\partial_{\nu}\log f)\left(
\hat{e}^{\nu \hat{b}}\hat{e}^{\mu\hat{a}}
-\hat{e}^{\nu \hat{a}}\hat{e}^{\mu\hat{b}}\right)
+\hat{\omega}_{\mu}{}^{\hat{a}\hat{b}},
\label{eq:omegamu}
\end{equation}
where the spin connection $\hat{\omega}_{\mu}{}^{\hat{a}\hat{b}}$
with the hat denotes the one for the vielbein \eqref{eq:hate}
for AdS$_{p+2}\times S^{8-p}$. 

Since the action \eqref{eq:fermionaction} is already quadratic
in the fermions, we just have to substitute the classical 
solution \eqref{eq:classical}, which will be called
$x^{\mu}_{(0)}$, into the bosonic fields $x^{\mu}$. 
The following combination often appears in the action,
\begin{align}
\partial_{\tau} x_{(0)}^{\mu}\Gamma_{\mu}
=cf(\bar{r})\left(
\cosh\rho_0\Gamma_{\hat{t}}+\sinh\rho_0\Gamma_{\hat{\psi}}
+\Gamma_{\hat{\tilde{\psi}}}\right)
=cf(\bar{r})\left(\hat{\Gamma}{}_{\hat{t}}
+\Gamma_{\hat{\tilde{\psi}}}\right).
\end{align}
In the last expression, we have defined the gamma matrix 
$\hat{\Gamma}{}_{\hat{t}}$ by the following transformation
from $\Gamma{}_{\hat{t}}$
and $\Gamma{}_{\hat{\psi}}$, 
\begin{align}
\begin{pmatrix} 
\hat{\Gamma}{}_{\hat{t}}\\ \hat{\Gamma}{}_{\hat{\psi}}
\end{pmatrix}
=
\begin{pmatrix}
\cosh\rho_0&\sinh\rho_0\\
\sinh\rho_0&\cosh\rho_0
\end{pmatrix}
\begin{pmatrix} 
\Gamma_{\hat{t}}\\ \Gamma_{\hat{\psi}}
\end{pmatrix}.
\label{eq:boost}
\end{align}
This is a Lorentz boost in the $\hat{t}$-$\hat{\psi}$ plane,
and the gamma matrices with the hat 
satisfy the same anti-commutation
relations as the original ones. The difference with the
$S=0$ case studied in \cite{AsanoSekino} is that we have 
$\hat{\Gamma}{}_{\hat{t}}$ in place of 
$\Gamma{}_{\hat{t}}$.

We take the following representation of the 10-dimensional
gamma matrices. For the $\hat{t}$ and $\hat{\tilde{\psi}}$ 
directions, we take
\begin{align}
\hat{\Gamma}{}_{\hat{t}} =\Gamma_{0}= 
\begin{pmatrix}0 & 1\cr -1& 0\end{pmatrix},\quad 
\Gamma_{\hat{\tilde{\psi}}} = \begin{pmatrix}0 & 1\cr 1& 0\end{pmatrix}.
\end{align}
where each block is a 16$\times$16 matrix.
For gamma matrices in the 
remaining 8 directions collectively denoted by $\Gamma_{\hat{i}}$
(with the one for the $\hat{\psi}$ direction
taken to be $\hat{\Gamma}{}_{\hat{\psi}}$ defined 
in \eqref{eq:boost}), we take 
\begin{align}
\Gamma_{\hat{i}} = \begin{pmatrix}\gamma_i & 0\cr 0& -\gamma_i\end{pmatrix}\, .
\end{align}
In this representation, 
\begin{align}
\Gamma_{11}=\begin{pmatrix}\gamma_9&0\cr 0& -\gamma_9\end{pmatrix}\,
\label{eq:Gamma11}
\end{align}
with $\gamma_9=\prod_{i=1}^8 \gamma_i$. 
We decompose $\Theta$ as
\begin{equation}
\Theta=\begin{pmatrix}\hat\theta \cr \theta\end{pmatrix}\, .
\label{eq:Theta}
\end{equation}
The action contains the following factor, which is proportional
to a projection operator $\hat{\Gamma}_{+}\equiv 
\Gamma_{0}(\hat{\Gamma}_{\hat{t}}+\Gamma_{\hat{\tilde{\psi}}})/2$,
\begin{equation}
\Gamma_{0}\partial_{\tau} x_{(0)}^{\mu}\Gamma_{\mu}
=-{2cLf(\bar{r})}\hat\Gamma_{+}
=-{2cLf(\bar{r})}\begin{pmatrix}0 & 0\cr 0& 1\end{pmatrix}.
\label{eq:project}
\end{equation}
This projects $\Theta$ on to the lower component,
so only half the degrees of freedom $\theta$ appear
in the action. The other half $\hat{\theta}$ is 
unphysical, and can be gauged away by $\kappa$-symmetry 
of the Green-Schwarz action~\cite{Cvetic, AsanoSekino}.

To rewrite the kinetic term (which depend on the covariant 
derivative $\nabla_\tau$), we make the following field
redefinition,
\begin{equation}
\Theta^{\rm (old)}=\left({Lf(\bar{r})\over 2c}\right)^{1\over 2}
e^{-{\tau\over 4}\partial_{\tau}x^{\mu}_{(0)}
\hat{\omega}_{\mu}{}^{\hat{a}\hat{b}}
\Gamma_{\hat{a}\hat{b}}}\Theta^{\rm (new)}.
\label{eq:redef}
\end{equation}
We chose the first factor on the right hand side so that 
the $f(\bar{r})$-dependent factor in the action disappears. 
At the same time, the $\tau$ derivative of this factor
cancels the first term in the spin connection
\eqref{eq:omegamu}. 
(This corresponds to the fact that the massless
spinor is Weyl invariant). We chose the second factor
so that the contribution to the covariant derivative 
which depends on the spin connection 
$\hat{\omega}{}_{\mu}{}^{\hat{a}\hat{b}}$ (which is 
constant when evaluated with the classical 
solution \eqref{eq:classical}) is canceled. 
In terms of the redefined (new) field, the kinetic term
becomes
\begin{align}
I_{\rm f,kin}
=+{i\over 2\pi}\int d\tau \Theta^{T}
\hat{\Gamma}_{+}\partial_{\tau}\Theta.
\label{eq:Ikin}
\end{align}

The mass term of the fermion comes from the ($p+2$)-form
field strength. We first note~\cite{AsanoSekino}
\begin{equation}
 \Omega= {7-p\over 8} \, L^{-1}f^{-1}(\bar{r})\Gamma_{(p)},
\end{equation}
where 
\begin{equation}
 \Gamma_{(p=0)}=\Gamma_{11} \Gamma^{\hat{s}\hat{x}_1},\quad
 \Gamma_{(p=2)}=-\Gamma^{\hat{s}\hat{x}_1 \hat{x}_2 \hat{x}_3},\quad
 \Gamma_{(p=4)}=-\Gamma_{11} \Gamma^{\hat{s}\hat{x}_1 \cdots \hat{x}_5}.
\end{equation}
Then, the $\Omega$-dependent term of the action becomes
\begin{align}
I_{{\rm f},\Omega}&=+{i\over 2\pi}\int d\tau \Theta^{\rm (old)}
{}^{T} \eta^{-1}\left(\Gamma_{0}
\partial_{\tau} x^{\mu}_{(0)}\Gamma_{\mu}\right)
\Omega\Gamma_{0}
\left(\Gamma_{0}\partial_{\tau} x^{\nu}_{(0)}
\Gamma_{\nu}\right)\Theta^{\rm (old)}
\nonumber\\
&=+{i\over 2\pi}\int d\tau \Theta^{T}
\hat{\Gamma}_{+}\left({7-p\over 2(5-p)}\right)
\Gamma_{(p)}\Gamma_{0}\hat{\Gamma}_{+}\Theta,
\label{eq:IOmega}
\end{align}
where the field $\Theta$ in the last line is the new field
defined in \eqref{eq:redef}. 
The $(p+2)$-form field strength enters in the action
through $f(\bar{r}) \Omega $. Since this is a constant,
it is not affected by the classical solution of the
bosonic field, and $I_{{\rm f},\Omega}$ takes 
the same value as in the $S=0$ case obtained in~\cite{AsanoSekino}.

Finally, by combining the kinetic and mass terms,
and writing the action 
in terms of the 16-component field $\theta$, we 
get
\begin{equation}
I_{\rm f}= - {i\over 2\pi}\int d\tau 
\theta^{T} 
\left(\partial_{\tau} \theta
+ m_{{\rm f}(p)} \gamma_{(p)}\theta\right)
\end{equation}
where 
\begin{equation}
m_{{\rm f}(p)}={7-p\over 2(5-p)}  
\label{eq:mfp}
\end{equation}
and 
\begin{equation}
 \gamma_{(p=0)}= \gamma_{9}\gamma_{1},\quad
 \gamma_{(p=2)}= \gamma_{123},\quad
 \gamma_{(p=4)}= -\gamma_{9}\gamma_{12345}.
\end{equation}
The Euclidean action is obtained from \eqref{eq:mfp} by
the standard procedure
$\tau\to i\tau_{\rm E}$, and multiplying the action 
by $-i$. The quantization of fermion described by this 
action is straightforward (see \cite{AsanoSekino}).
The action for the type IIB theory can be obtained
in exactly the same manner as above, 
and we find that the mass is also 
given by \eqref{eq:mfp} (with odd $p$)~\cite{AsanoSekino}.


\begin{thebibliography}{10}

\bibitem{Holography}
Leonard Susskind.
\newblock {The World as a hologram}.
\newblock {\em J. Math. Phys.}, 36:6377--6396, 1995.

\bibitem{tHooft}
Christopher~R. Stephens, Gerard 't~Hooft, and Bernard~F. Whiting.
\newblock {Black hole evaporation without information loss}.
\newblock {\em Class. Quant. Grav.}, 11:621--648, 1994.

\bibitem{Maldacena}
Juan~Martin Maldacena.
\newblock {The Large N limit of superconformal field theories and
  supergravity}.
\newblock {\em Adv. Theor. Math. Phys.}, 2:231--252, 1998.

\bibitem{chaos1}
Stephen~H. Shenker and Douglas Stanford.
\newblock {Black holes and the butterfly effect}.
\newblock {\em JHEP}, 03:067, 2014.

\bibitem{RT1}
Shinsei Ryu and Tadashi Takayanagi.
\newblock {Holographic derivation of entanglement entropy from AdS/CFT}.
\newblock {\em Phys. Rev. Lett.}, 96:181602, 2006.

\bibitem{RT2}
Shinsei Ryu and Tadashi Takayanagi.
\newblock {Aspects of Holographic Entanglement Entropy}.
\newblock {\em JHEP}, 08:045, 2006.

\bibitem{chaos2}
Juan Maldacena, Stephen~H. Shenker, and Douglas Stanford.
\newblock {A bound on chaos}.
\newblock {\em JHEP}, 08:106, 2016.

\bibitem{SakaiSugimoto1}
Tadakatsu Sakai and Shigeki Sugimoto.
\newblock {Low energy hadron physics in holographic QCD}.
\newblock {\em Prog. Theor. Phys.}, 113:843--882, 2005.

\bibitem{SakaiSugimoto2}
Tadakatsu Sakai and Shigeki Sugimoto.
\newblock {More on a holographic dual of QCD}.
\newblock {\em Prog. Theor. Phys.}, 114:1083--1118, 2005.

\bibitem{Sekino}
Yasuhiro Sekino.
\newblock {Evidence for weak-coupling holography from the gauge/gravity
  correspondence for D$p$-branes}.
\newblock {\em PTEP}, 2020(2):021B01, 2020.

\bibitem{WittenAdS}
Edward Witten.
\newblock {Anti-de Sitter space, thermal phase transition, and confinement in
  gauge theories}.
\newblock {\em Adv. Theor. Math. Phys.}, 2:505--532, 1998.

\bibitem{GKP}
S.~S. Gubser, Igor~R. Klebanov, and Alexander~M. Polyakov.
\newblock {Gauge theory correlators from noncritical string theory}.
\newblock {\em Phys. Lett. B}, 428:105--114, 1998.

\bibitem{Ferrara}
Sergio Ferrara, Christian Fronsdal, and Alberto Zaffaroni.
\newblock {On N=8 supergravity on AdS(5) and N=4 superconformal Yang-Mills
  theory}.
\newblock {\em Nucl. Phys. B}, 532:153--162, 1998.

\bibitem{Semenoff}
J.~K. Erickson, G.~W. Semenoff, R.~J. Szabo, and K.~Zarembo.
\newblock {Static potential in N=4 supersymmetric Yang-Mills theory}.
\newblock {\em Phys. Rev. D}, 61:105006, 2000.

\bibitem{Eden}
Niklas Beisert, Burkhard Eden, and Matthias Staudacher.
\newblock {Transcendentality and Crossing}.
\newblock {\em J. Stat. Mech.}, 0701:P01021, 2007.

\bibitem{WittenThermal}
Edward Witten.
\newblock {Anti-de Sitter space and holography}.
\newblock {\em Adv. Theor. Math. Phys.}, 2:253--291, 1998.

\bibitem{PilchWarner}
D.~Z. Freedman, S.~S. Gubser, K.~Pilch, and N.~P. Warner.
\newblock {Renormalization group flows from holography supersymmetry and a c
  theorem}.
\newblock {\em Adv. Theor. Math. Phys.}, 3:363--417, 1999.

\bibitem{KlebanovStrassler}
Igor~R. Klebanov and Matthew~J. Strassler.
\newblock {Supergravity and a confining gauge theory: Duality cascades and chi
  SB resolution of naked singularities}.
\newblock {\em JHEP}, 08:052, 2000.

\bibitem{PolchinskiStrassler}
Joseph Polchinski and Matthew~J. Strassler.
\newblock {Hard scattering and gauge / string duality}.
\newblock {\em Phys. Rev. Lett.}, 88:031601, 2002.

\bibitem{LiuReview}
Jorge Casalderrey-Solana, Hong Liu, David Mateos, Krishna Rajagopal, and
  Urs~Achim Wiedemann.
\newblock {\em {Gauge/String Duality, Hot QCD and Heavy Ion Collisions}}.
\newblock Cambridge University Press, 2014.

\bibitem{HartnollReview}
Sean~A. Hartnoll, Andrew Lucas, and Subir Sachdev.
\newblock {Holographic quantum matter}.
\newblock 12 2016.

\bibitem{IMSY}
Nissan Itzhaki, Juan~Martin Maldacena, Jacob Sonnenschein, and Shimon
  Yankielowicz.
\newblock {Supergravity and the large N limit of theories with sixteen
  supercharges}.
\newblock {\em Phys. Rev. D}, 58:046004, 1998.

\bibitem{SekinoYoneya}
Yasuhiro Sekino and Tamiaki Yoneya.
\newblock {Generalized AdS / CFT correspondence for matrix theory in the large
  N limit}.
\newblock {\em Nucl. Phys. B}, 570:174--206, 2000.

\bibitem{SekinoSupercurrents}
Yasuhiro Sekino.
\newblock {Supercurrents in matrix theory and the generalized AdS / CFT
  correspondence}.
\newblock {\em Nucl. Phys. B}, 602:147--171, 2001.

\bibitem{SkenderisDp}
Ingmar Kanitscheider, Kostas Skenderis, and Marika Taylor.
\newblock {Precision holography for non-conformal branes}.
\newblock {\em JHEP}, 09:094, 2008.

\bibitem{HNSY1}
Masanori Hanada, Jun Nishimura, Yasuhiro Sekino, and Tamiaki Yoneya.
\newblock {Monte Carlo studies of Matrix theory correlation functions}.
\newblock {\em Phys. Rev. Lett.}, 104:151601, 2010.

\bibitem{HNSY2}
Masanori Hanada, Jun Nishimura, Yasuhiro Sekino, and Tamiaki Yoneya.
\newblock {Direct test of the gauge-gravity correspondence for Matrix theory
  correlation functions}.
\newblock {\em JHEP}, 12:020, 2011.

\bibitem{MaldacenaOoguri1}
Juan~Martin Maldacena and Hirosi Ooguri.
\newblock {Strings in AdS(3) and SL(2,R) WZW model 1.: The Spectrum}.
\newblock {\em J. Math. Phys.}, 42:2929--2960, 2001.

\bibitem{MaldacenaOoguri2}
Juan~Martin Maldacena, Hirosi Ooguri, and John Son.
\newblock {Strings in AdS(3) and the SL(2,R) WZW model. Part 2. Euclidean black
  hole}.
\newblock {\em J. Math. Phys.}, 42:2961--2977, 2001.

\bibitem{MaldacenaOoguri3}
Juan~Martin Maldacena and Hirosi Ooguri.
\newblock {Strings in AdS(3) and the SL(2,R) WZW model. Part 3. Correlation
  functions}.
\newblock {\em Phys. Rev. D}, 65:106006, 2002.

\bibitem{D1D5}
Justin~R. David, Gautam Mandal, and Spenta~R. Wadia.
\newblock {Microscopic formulation of black holes in string theory}.
\newblock {\em Phys. Rept.}, 369:549--686, 2002.

\bibitem{BMN}
David~Eliecer Berenstein, Juan~Martin Maldacena, and Horatiu~Stefan Nastase.
\newblock {Strings in flat space and pp waves from N=4 superYang-Mills}.
\newblock {\em JHEP}, 04:013, 2002.

\bibitem{GKP2}
S.~S. Gubser, I.~R. Klebanov, and Alexander~M. Polyakov.
\newblock {A Semiclassical limit of the gauge / string correspondence}.
\newblock {\em Nucl. Phys. B}, 636:99--114, 2002.

\bibitem{DSY}
Suguru Dobashi, Hidehiko Shimada, and Tamiaki Yoneya.
\newblock {Holographic reformulation of string theory on AdS(5) x S**5
  background in the PP wave limit}.
\newblock {\em Nucl. Phys. B}, 665:94--128, 2003.

\bibitem{DobashiYoneya1}
Suguru Dobashi and Tamiaki Yoneya.
\newblock {Resolving the holography in the plane-wave limit of AdS/CFT
  correspondence}.
\newblock {\em Nucl. Phys. B}, 711:3--53, 2005.

\bibitem{Dobashi}
Suguru Dobashi.
\newblock {Impurity Non-Preserving 3-Point Correlators of BMN Operators from
  PP-Wave Holography. II. Fermionic Excitations}.
\newblock {\em Nucl. Phys. B}, 756:171--206, 2006.

\bibitem{ASY}
Masako Asano, Yasuhiro Sekino, and Tamiaki Yoneya.
\newblock {PP wave holography for Dp-brane backgrounds}.
\newblock {\em Nucl. Phys. B}, 678:197--232, 2004.

\bibitem{AsanoSekino}
Masako Asano and Yasuhiro Sekino.
\newblock {Large N limit of SYM theories with 16 supercharges from superstrings
  on Dp-brane backgrounds}.
\newblock {\em Nucl. Phys. B}, 705:33--59, 2005.

\bibitem{Minwalla}
Chi-Ming Chang, Shiraz Minwalla, Tarun Sharma, and Xi~Yin.
\newblock {ABJ Triality: from Higher Spin Fields to Strings}.
\newblock {\em J. Phys. A}, 46:214009, 2013.

\bibitem{Gaberdiel}
Matthias~R. Gaberdiel and Rajesh Gopakumar.
\newblock {String Dual to Free $N=4$ Supersymmetric Yang-Mills Theory}.
\newblock {\em Phys. Rev. Lett.}, 127(13):131601, 2021.

\bibitem{SusskindWitten}
Leonard Susskind and Edward Witten.
\newblock {The Holographic bound in anti-de Sitter space}.
\newblock hep-th/9805114, 1998.

\bibitem{Banks}
Tom Banks and Michael~B. Green.
\newblock {Nonperturbative effects in AdS in five-dimensions x S**5 string
  theory and d = 4 SUSY Yang-Mills}.
\newblock {\em JHEP}, 05:002, 1998.

\bibitem{Kallosh}
Renata Kallosh and Arvind Rajaraman.
\newblock {Vacua of M theory and string theory}.
\newblock {\em Phys. Rev. D}, 58:125003, 1998.

\bibitem{Metsaev1}
R.~R. Metsaev.
\newblock {Type IIB Green-Schwarz superstring in plane wave Ramond-Ramond
  background}.
\newblock {\em Nucl. Phys. B}, 625:70--96, 2002.

\bibitem{Metsaev2}
R.~R. Metsaev and Arkady~A. Tseytlin.
\newblock {Exactly solvable model of superstring in Ramond-Ramond plane wave
  background}.
\newblock {\em Phys. Rev. D}, 65:126004, 2002.

\bibitem{Cvetic}
Mirjam Cvetic, H.~Lu, C.~N. Pope, and K.~S. Stelle.
\newblock {Linearly realised world sheet supersymmetry in pp wave background}.
\newblock {\em Nucl. Phys. B}, 662:89--119, 2003.

\bibitem{Jevicki1}
Antal Jevicki and Tamiaki Yoneya.
\newblock {Space-time uncertainty principle and conformal symmetry in D
  particle dynamics}.
\newblock {\em Nucl. Phys. B}, 535:335--348, 1998.

\bibitem{Jevicki2}
Antal Jevicki, Yoichi Kazama, and Tamiaki Yoneya.
\newblock {Quantum metamorphosis of conformal transformation in D3-brane
  Yang-Mills theory}.
\newblock {\em Phys. Rev. Lett.}, 81:5072--5075, 1998.

\bibitem{Jevicki3}
Antal Jevicki, Yoichi Kazama, and Tamiaki Yoneya.
\newblock {Generalized conformal symmetry in D-brane matrix models}.
\newblock {\em Phys. Rev. D}, 59:066001, 1999.

\bibitem{Asano}
Masako Asano.
\newblock {Stringy effect of the holographic correspondence for dp-brane
  backgrounds}.
\newblock {\em JHEP}, 12:029, 2004.

\bibitem{Tseytlin1}
S.~Frolov and Arkady~A. Tseytlin.
\newblock {Semiclassical quantization of rotating superstring in AdS(5) x
  S**5}.
\newblock {\em JHEP}, 06:007, 2002.

\bibitem{Tseytlin2}
S.~Frolov and Arkady~A. Tseytlin.
\newblock {Multispin string solutions in AdS(5) x S**5}.
\newblock {\em Nucl. Phys. B}, 668:77--110, 2003.

\bibitem{Tseytlin3}
S.~Frolov and Arkady~A. Tseytlin.
\newblock {Quantizing three spin string solution in AdS(5) x S**5}.
\newblock {\em JHEP}, 07:016, 2003.

\bibitem{WIP}
Tomotaka Kitamura, Shoichiro Miyashita, and Yasuhiro Sekino.
\newblock work in progress.

\end{thebibliography}
\end{document}